 \documentclass[preprint2,longabstract]{aastex}

\usepackage{natbib}

\newcommand{\msun}{\,{M}_{\odot}}
\newcommand{\lsun}{\,{L}_{\odot} }

\newcommand{\esc}{\,{\rm erg\,s^{-1}\,cm^{-2}}}
\newcommand{\kms}{\,{\rm km\,s^{-1}}}      
\newcommand{\mum}{\,\mu{\rm m}}

\newcommand{\solar}{L$_{\odot}$\ }
\newcommand{\solm}{M$_{\odot}$}

\slugcomment{}

\shorttitle{Monitoring the DSO}
\shortauthors{Valencia-S. et al.}

\begin{document}


\title{
Monitoring the Dusty S-Cluster Object (DSO/G2) on its Orbit towards the Galactic Center Black Hole
}

\author{ M. Valencia-S.\altaffilmark{1}, 
 A. Eckart\altaffilmark{1,2}, 
 M. Zaja\v{c}ek\altaffilmark{1,2,3},
 F. Peissker\altaffilmark{1},
 M. Parsa\altaffilmark{1},
 N. Grosso\altaffilmark{4},
 E. Mossoux\altaffilmark{4},
 D. Porquet\altaffilmark{4},
 B. Jalali\altaffilmark{1},
 V. Karas\altaffilmark{3},
 S. Yazici\altaffilmark{1},
 B. Shahzamanian\altaffilmark{1},
 N. Sabha\altaffilmark{1},
 R. Saalfeld\altaffilmark{1},
 S. Smajic\altaffilmark{1}, 
 R. Grellmann\altaffilmark{1}, 
 L. Moser\altaffilmark{1},
 M. Horrobin\altaffilmark{1},
 A. Borkar\altaffilmark{1}, 
 M. Garc\'{i}a-Mar\'{i}n\altaffilmark{1},
 M. Dov\v{c}iak\altaffilmark{3}, 
 D. Kunneriath\altaffilmark{3}, 
 G. D. Karssen\altaffilmark{1},
 M. Bursa\altaffilmark{3},
 C. Straubmeier\altaffilmark{1},
and
 H. Bushouse\altaffilmark{5}
}

\altaffiltext{1}{I. Physikalisches Institut der Universit\"at zu K\"oln, Z\"ulpicher Str. 77, D-50937 K\"oln, Germany; mvalencias@ph1.uni-koeln.de}
\altaffiltext{2}{Max-Planck-Institut f\"ur Radioastronomie, Auf dem H\"ugel 69, D-53121 Bonn, Germany}
\altaffiltext{3}{Astronomical Institute of the Academy of Sciences Prague, Bo\v{c}n\'{i} II 1401/1a, CZ-141 31 Praha 4, Czech Republic}
\altaffiltext{4}{Observatoire Astronomique de Strasbourg, Universit\'e de Strasbourg, CNRS, UMR 7550, 11 rue de l'Universit\'e, F-67000 Strasbourg, France}
\altaffiltext{5}{Space Telescope Science Institute, Baltimore, MD 21218, USA}


\begin{abstract}
We analyse and report in detail new near-infrared (1.45 - 2.45 microns) 
observations of the Dusty S-cluster Object (DSO/G2) during its approach to the black hole at the center of the 
Galaxy that were carried out with ESO VLT/SINFONI between February and September 2014.
Before May 2014 we detect spatially compact Br$\gamma$ and Pa$\alpha$ line emission from the DSO at about 40mas east of SgrA*. 
The velocity of the source, measured from the red-shifted emission, is 2700$\pm$60 km/s. 
No blue-shifted emission above the noise level is detected at the position of SgrA* or upstream the presumed orbit. 
After May we find spatially compact Br$\gamma$ blue-shifted line emission from the DSO at about 30mas west of SgrA*
at a velocity of $-$3320$\pm$60 km/s and no indication for significant red-shifted emission.
We do not detect any significant extension of velocity gradient across the source.
We find a Br$\gamma$-line full width at half maximum of 
50$\pm$10~\AA ~before and 15$\pm$10~\AA ~after the peribothron transit,
i.e. no significant line broadening with respect to last year is observed. 
Br$\gamma$ line maps show that the bulk of the line emission originates from a region of less than 20~mas diameter.
This is consistent with a very compact source on an elliptical orbit with a peribothron time passage in 2014.39$\pm$0.14.
For the moment, the flaring activity of the black hole in the near-infrared regime has not shown any 
statistically significant increment.
Increased accretion activity of SgrA* may still be upcoming.
We discuss details of a source model according to which the DSO is rather a young accreting star than a coreless gas and dust cloud.
\end{abstract}





\section{Introduction}
\label{sec:introduction}

Recently the Galactic center region has attracted a lot of attention
due to the fact that a dusty object has been detected \citep{gil12, gil13a, eck13} 
that is approaching the central super massive black hole associated with 
the radio source SgrA*.
Due to its infrared excess and as indicated through nomenclature (G2) 
\vspace{4.5cm}
\noindent
it has been speculated that the source 
consists to a dominant fraction of gas and dust \citep{gil12, gil13a, pfu14b}. 
By now the object is expected to have passed through its peribothron and
tidal disruption as well as intense accretion events have been predicted.
\citet{eck13} show a possible spectral decomposition
of this source using the M-band measurement by \citet{gil12}.
Depending on the relative stellar and dust flux density contributions the
M-band measurement 
is consistent with a dust temperature of 450~K and an integrated 
luminosity of up to $\sim 10\lsun$.
This allows for a substantial stellar contribution in mass and reddened stellar luminosity.
A stellar nature is also favored by many other authors 
\citep[see also][]{mur12, sco13, bal13, phi13, zajacek14}.
We will therefore refer to it in the following as a Dusty S-cluster Object, DSO \citep{eck13}.
Hence, although the Br$\gamma$ line emission may be dominated by optically thin emission,
a contribution from more compact optically thick regions cannot be excluded. 
Also, it is uncertain how large the extinction towards the center of the gas cloud really is. 
Therefore, the total mass of the object is very uncertain but is presumably less than that of a
typical member of the high velocity S-star cluster 
\citep[i.e. $\lesssim 20\msun$;][]{ghez03, eis05, mar06}.
The compactness of the DSO is also supported by the recent L-band detection close to 
peribothron \citep{ghez14, witzel14}.

\citet{gil13a, gil13b} and \citet{pfu14b} 
report that the Br$\gamma$ luminosity of the DSO has remained 
constant over the entire time range covered by spectroscopy from 2004 to 2013. 
Figs.~1 and 5 in \citet{pfu14b} show 
that in their April 2014 data set the 
blue line emission is approximately as spatially compact as the red side and has a 
significantly stronger peak emission than the red line emission. 
Their derived integrated Br$\gamma$ luminosities for the blue side is 
about 1.14 times brighter than the red side \citep[section 3.2 in][]{pfu14b}.

During the past year we have obtained a substantial, independent imaging spectroscopy
data set using SINFONI at the ESO VLT. In addition we have re-reduced a large 
number of data sets available from the ESO archive and have used our own and published
positional data to re-estimate the orbit of the DSO.
Here we present the results of this detailed investigation.
The paper is organized in the following way:
In sections 2 and 3 we
present the observations and data reduction, including the analysis of the spectral line properties
of the DSO. In section 4 we discuss the results in-
cluding the orbit (section 4.1), the tidal interaction
of the DSO with SgrA* (section 4.2) and the ambient 
medium (section 4.3), consequences for the
flare activity (section 4.4), and the interpretation
of the DSO as a possible pre-main sequence star
(section 4.5). After discussing the origin and fate
of the DSO in section 5 we summarize and conclude
in section 6.

\section{Observations and data reduction}
\label{sec:obs}

Here we present the data sets we are using in the study of the DSO.
The procedures for data reduction and data quality selection are also described.
We report mainly on the observations\footnote{ESO programs 092.B-0009 (PI: A. Eckart), 
093.B-0092 (PI: A. Eckart) and 092.B-0920 (PI: N. Grosso)}
we conducted from February to September 2014.
We use earlier archive data to discuss general properties like the DSO orbit.

\subsection{The 2014 Data Set}
\label{sec:dataset}

We performed NIR integral field observations of the Galactic center using
SINFONI at the VLT in Chile \citep{eis03,bon04}. 
The instrument is an image-slicer integral field unit 
fed by an adaptive optics (AO) module.
The AO system uses an optical wavefront sensor that was locked on 
a bright star $15.54''$ north and $8.85''$ east of Sgr~A*.
We employed H+K grating that covers the $1.45 - 2.45 \mum$ range with 
a spectral resolution of $R \sim1500$ (i.e., approximately $200\kms$ at $2.16\mum$).
The $0.8''\times0.8''$ field of view was jittered around the position of the star S2, 
in such a way that the star remained within the upper half zone of the detector.
This was done in order to avoid a region with possible non-linear behavior of the detector.
Observations of different B- and G-type stars were taken to obtain independent telluric
templates.

The Galactic center region was observed in intervals of ${\rm 400\,s}$ or ${\rm 600\,s}$,
followed or preceded by time slots of equal length on a dark cloud  $5'36''$ north
and $12'45''$ west of Sgr~A*.
The integration time were chosen of that length in oder to be able to also monitor the 
flux density of SgrA* for time intervals of typical flare lengths and to provide a
higher flexibility in data selection to optimize the quality of the data.
Although this observational strategy reduces the effective integration time
on source to about a third when compared with parsed sky observations
at a rate of about once per hour,
it ensures a better control of the noise in the frames.
Because of the variable weather conditions, the point-spread function (PSF) changed
along the observing nights.
The quality of individual exposures was judged based on the PSF at the moment of the observation,
 as measured from the shape of the stars in the field of view.
For the analysis presented here, we have created two final data cubes,
one from the combination of the best quality exposures, and another including also
medium quality data, as described below.
Table~\ref{tab1} shows the list of the observing dates, including the amount of exposures
that fulfilled the selection criteria. 
Note that (both for pre- and post-peribothron) our observations are bracketing and preceeding those 
presented by \citet{pfu14b}.

\begin{deluxetable}{ccccccc}
\tabletypesize{\scriptsize}
\tablecaption{Summary of the Galactic Center Observations \label{tab1}}
\tablewidth{0pt}
\tablehead{
\colhead{Date} & \colhead{Start time} & \colhead{End time} & \multicolumn{3}{c}{Number of on-source exposures}                    & \colhead{Exp. Time} \\ \cline{4-6}
               & \colhead{}           & \colhead{}         & \colhead{Total} & \colhead{medium quality} & \colhead{high quality} & \colhead{} \\
 (YYYY.MM.DD)  & \colhead{(UT)}       & \colhead{(UT)}     & \colhead{}      & \colhead{}                       & \colhead{}      &   \colhead{(s)}               
}
\startdata
2014.02.28 & 08:34:58  & 09:54:37  &  7 &   0  &   0  &    400  \\
2014.03.01 & 08:00:14  & 10:17:59  & 12 &   0  &   0  &    400  \\
2014.03.02 & 07:49:06  & 08:18:54  &  3 &   0  &   0  &    400  \\
2014.03.11 & 08:03:55  & 10:03:28  & 11 &   5  &   8  &    400  \\
2014.03.12 & 07:44:35  & 10:07:45  & 13 &   5  &   9  &    400  \\
2014.03.26 & 06:43:05  & 09:58:12  & 11 &   8  &   8  &    600  \\
2014.03.27 & 06:32:50  & 10:04:12  & 18 &   1  &   7  &    400  \\
2014.04.02 & 06:31:39  & 09:53:52  & 18 &   0  &   5  &    400  \\
2014.04.03 & 06:20:46  & 09:45:02  & 18 &  14  &  17  &    400  \\
2014.04.04 & 05:58:19  & 09:47:58  & 21 &  14  &  17  &    400  \\
2014.04.06 & 07:51:42  & 08:43:15  &  5 &   4  &   1  &    400  \\ \hline
2014.06.09 & 04:48:49  & 09:51:47  & 14 &  14  &   0  &    400  \\
2014.06.10 & 04:54:21  & 09:49:49  &  5 &   5  &   0  &    400  \\
2014.08.25 & 23:57:46  & 04:34:49  &  4 &   4  &   0  &    400  \\
2014.09.07 & 00:11:08  & 04:20:07  &  2 &   2  &   0  &    400  \\
\enddata
\tablecomments{~List of Start and End times, number and quality of exposures.
ESO program 092.B-0009 for 
2014.02.28, 2014.03.01, 2014.03.02, 2014.03.26, 2014.03.27, and 2014.04.06,
ESO program 092.B-0920 for 
2014.03.11, 2014.03.12, 2014.04.02,2014.04.03, and 2014.04.04
and ESO program 093.B-0092 for
2014.06.09, 2014.06.10, 2014.08.25, 2014.09.07.
Pre- and post-peribothron measurements are separated by a horizontal line.
}
\end{deluxetable}

\subsection{Calibration}  
\label{calibration}  

In the data reduction process, we first followed the SINFONI reduction manually to correct for the
bad lines created by the data processing at the detector level. We used the suggested IDL procedure, 
adjusting the identification threshold (2 times the background noise $\sigma_{background}$)  whenever necessary. 
A first cosmic ray correction to the sky and target files was performed using the algorithm developed by  \citet{pych04}. 
The random pattern introduced by some detector amplifiers was detected and removed in science 
and calibration files following the algorithms proposed by \citet{smajic14}. 
Then we used the SINFONI pipeline for the standard reduction steps (like flat fielding and bad pixel corrections) 
and wavelength calibration. We obtained one data cube for each on-source exposure. 

DPUSER routines \citep[Thomas Ott, MPE Garching; see also][]{eck91} were used to flag remaining bad pixels and cosmic rays 
on the plane of the slitlets in the detector ($x$-$z$, which corresponds to dec.-wavelength),  in each data cube. 
The combined effects of the atmospheric refraction 
 were appreciable as a spatial displacement of the stars by a couple of pixels 
when going from short to long wavelengths.   
Fixing the position of the center of a bright source at a particular wavelength and 
making a spatial sub-pixel shift at all other wavelengths could correct the problem, but 
the shape of the resulting spectrum in each pixel would depend on the interpolation algorithm. 
Therefore,   to preserve the integrity of the spectrum in the narrow spectral regions 
 where  emission lines are present, the spatial image shift was done  in steps of $0.045\mum$.

A 2D Gaussian, fitted to the bright star S2, was taken as an indication of the PSF. 
Cubes where the FWHM of the fitted Gaussian is less than 83\,mas (or 6.65 pixels) were categorized
as  best quality cubes, while those with FWHM values between 83 and 96\,mas (or 7.65 pixels) were classified as  
medium  quality  cubes.
The combination of the selected data cubes was done by averaging every 
spatial and spectral pixel after a proper alignment of the images.
The combination of the 63 best quality cubes produced a final data cube with a total of 7.2\,h 
 on-source integration time.  
When including the 30 medium quality exposures, the resulting data cube covers a total of 10.8\,h 
of integration time on-source. This second data cube was used to evaluate the 
effects of the data quality in the signal-to-noise of the measured quantities, and unless it is 
specifically mentioned in the text, all measurements and plots are derived from the higher quality data cube.

Flux calibration was done using aperture photometry on a deconvolved $K$-band image 
created from the final data cube. 
The deconvolution was performed using the Lucy-Richardson algorithm incorporated in DPUSER,
while the PSF was estimated using the IDL based StarFinder routine \citep{dio00}. 
We used as calibration stars S2 ($K_{\rm s}=14.1$), S4 ($K_{\rm s}=14.6$), S10 ($K_{\rm s}=14.1$), 
and S12 ($K_{\rm s}=15.2$), 
and adopted the $K$-band extinction correction  $m_{A_K}=2.46$ of 
\citet{scho10}; 
\citep[see also][for the flux estimation]{witzel12}.

The NIR spectrum of the inner $\sim 0.5$ arcseconds around SgrA* is dominated by 
the stellar continuum of 
hundreds of stars fainter than $K_{\rm s}=18 \,{\rm mag}$ that are part of the central cluster, and 
that are unresolved with the current instrumentation \citep{sabha12}. 
Several absorption features from the stellar atmospheres can be recognized in the {\it HK}-band data. 
Line emission of ionized species (hydrogen and helium) at the position of the S-stars, 
and all across the field is also substantial. 
We refer to the aggregate of all these components as background, 
and show how it affects the detection of the faint emission of the DSO.

\section{Results}  
\label{sec:results}  

In summary, we find that both the line shape and line intensity in 2014 is very similar 
to that of the previous years.
Before May 2014 we find no blue line emission from hydrogen or helium above the noise level.
The red-line center has shifted to a higher velocity of about 2700 $\pm$ 60 km/s about 40~mas east of SgrA*.
In addition we measure the Pa$\alpha$ $\lambda$1.875 $\mu$m line between atmospheric absorption bands 
but find that the sky subtracted HeI ($2.05\,\mu$m) is very weak 
i.e. less than a fifth of the Br$\gamma$ line emission. 
In June 2014 our data does not allow use to detect the HeI or Pa$\alpha$ line emission.
Peribothron happened in May 2014 and after that we see a in June 2014 
blue-shifted Br$\gamma$ line about 16 west of SgrA* at -3320~km/s and no red line emission.
In June 2014 we find no blue line emission helium, Br$\gamma$ or Pa$\alpha$ above the noise level.

\subsection{Red-shifted pre-peribothron lines}
\label{subsec:red}

In 2014, the Br$\gamma$ line emission from the DSO has shifted to a spectral region 
where the emission and absorption features of the surroundings are very prominent.  
Moreover, the Pa$\alpha$ and the possible \ion{He}{1} lines lie in a wavelength range 
where the atmospheric absorption plays a main role.  
Therefore, a proper background estimation and subtraction, as well as an adequate 
fitting and correction of the tellurics, are critical to detect and measure any emission 
from the DSO. 
The latter is expected to change from one day to the next, and also during one 
observing night if the weather conditions are unstable, but it is approximately the same 
across the field of view. The former, on the other hand, varies strongly across the field 
and significantly in periods of about six months, because of the high stellar proper motions.
This means that, while the telluric absorption can be, in principle, fitted and corrected 
using extra observations of stars or sources in the field, the exact shape of the 
background spectrum at the position of the DSO cannot be known, but only approximated 
by using apertures in the field.  
A carefully calculated approximation to the background leads to a reliable estimation 
of the spectral properties of the source's line emission
as well as the the spatial position and extension of the DSO emission, 

\subsubsection{The Br$\gamma$ emission line}
\label{subsubsec:brg}

\begin{figure*}[p!]
\includegraphics[width=\textwidth, bb = 20 160 580 740, clip]{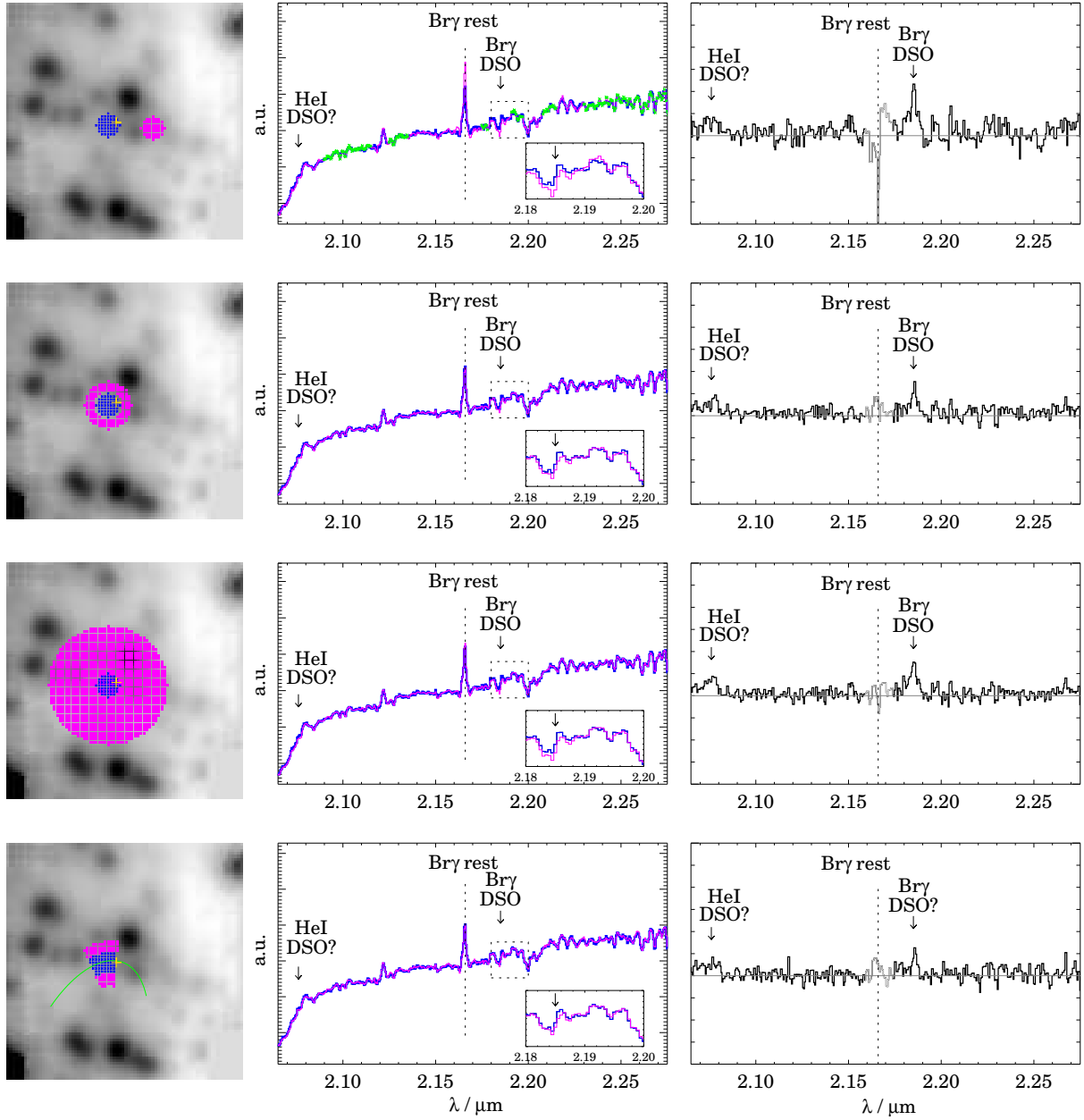}
\caption{
Br$\gamma$ red-shifted emission of the DSO before May at 43mas east and 5mas south of SgrA*.
{\it Left panels:} $1 \times 1\,{\rm arcsec^{2}}$ SINFONI mosaic of the Galactic center region in February to April 2014.
DSO (blue crosses) and background (pink squares) apertures are shown. {\it Central panels:} Comparison between the DSO
(thick line) and the background (thin line) spectra. Arrows mark the expected location of the red-shifted \ion{He}{1} and
Br$\gamma$ lines. {\it Right panels:} Results of the subtraction of the background from the DSO spectrum.
The baseline is shown in gray. The vertical range of the plots corresponds to one unit in the middle panels.
See details in the text. See the electronic edition of the Journal for a color version of this figure.
\label{figdsored}}
\end{figure*}

Figure \ref{figdsored} shows the spectrum integrated over apertures of radius $0.05''$
at the position where the red-shifted Br$\gamma$ emission from the DSO is strongest. 
It also shows the background spectra constructed from four different surrounding regions, 
and the resulting line emission after subtracting them from the source spectrum. 
On the left-hand panels, the spatial pixels from which the spectra of the source and the background
have been extracted are marked with blue crosses and pink filled squares, respectively.
The SgrA* position is marked with the big cross.
The middle panels display the integrated spectrum at the DSO position (thick blue line)
in comparison with the background spectra (thin pink line).
The vertical dashed-line at $2.166\mum$ marks the spectral position of the zero-velocity Br$\gamma$ line.
Arrows at $2.076\mum$ and $2.185\mum$ indicate the approximate location of the DSO 
red-shifted \ion{He}{1} and Br$\gamma$ emission lines.
The inset panel corresponds to the dashed-line box, which is a zoom-in to the spectra in the $2.18-2.20\mum$ range.
The arrow in the inset panel marks again the position of the red-shifted Br$\gamma$ line
and highlights the importance of the background subtraction to recover the spectral properties of the emission.
Because of the change of the spectral slope across the field of view, the overall shape of the spectrum 
extracted from the background area has to be slightly modified to better fit the continuum in the DSO aperture.  
To do that, we divide it by a third-order polynomial fitted to the ratio between the source spectrum and the 
spectrum of the background aperture.
This is done using only the spectral windows marked with (green) crosses in the top panel of the middle column.
Then, the background was scaled to best match the continuum emission around the spectral location of 
the Br$\gamma$ red-shifted emission (i.e, at $2.173-2.183\mum$ and $2.195-2.220\mum$).
These spectra, with modified slope and scaled continuum, are used as an approximation of the
background emission at the DSO position.
They reproduce well most of the features in the source spectrum, as it can be seen in the middle panels.
Given that at wavelengths shorter than $\sim 2.08\mum$ the emission is highly absorbed by 
tellurics, we did not included this spectral region in the fit of the overall background-spectral shape,
but used only the selected spectral windows as described above.
For this reason, the background continuum in the source aperture at $\lambda \lesssim 2.08\mum$
can not be fitted properly and produces an excess of emission in the  background-subtracted DSO spectra 
that can be seen in the right-hand panels.
There, the large mismatch observed in the spectra in a range of $0.82\mum$ around 
the zero-velocity Br$\gamma$ line is due to strong variations of the ionized hydrogen emission 
in the central $r\, \sim 1''$ region.
Lines within the telluric absorption region are treated differently to improve their signal
strength, see section~\ref{subsubsec:otherlinered}.

The four examples shown in Fig.~\ref{figdsored} correspond to the cases when:
1) The background is created from an aperture of the same size
and shape as that of the source, and it is located just beside it.
2) Iris photometry is applied, i.e. the background aperture is a ring around the source aperture.
In this case, the inner radius was chosen to be $0.06''$ and the outer radius, $0.11''$.
3) An averaged background is created from a region of radius $0.25''$ at the source position that includes the DSO aperture.
4) The source emission is integrated in a segment of $0.075''$ width taken  along
the best-fit elliptical orbit with a length of $\sim0.10''$ (see section~\ref{subsec:orbit}). 
The background is integrated from the $0.048''$-width stripes above and below the source area.
In all cases the red-shifted Br$\gamma$ line is detected with signal-to-noise ratios
between 3.9 (in the second case) and 4.7 (in the third one).

Fitting a Gaussian to the line emission in each case, we find a rather robust determination 
of the line peak at $2.185\mum$, i.e., 
 $2700\kms$ in average with a variation of $60\kms$.
However, as it can be noticed from Fig.~\ref{figdsored}, the line profile changes depending on the subtracted background.
The FWHM of the Br$\gamma$ line, corrected for instrumental broadening, is 
$730\kms$ in the first and third cases, $560\kms$ in the second, and only $240\kms$ in the fourth one.
Averaging over a dozen background-subtracted source apertures the FWHM(Br$\gamma) \approx 720\pm 150\kms$,
i.e. the line width is 50$\pm$10~\AA.
The line flux changes, in general, by a factor of two due to the background subtraction.
In the first three examples shown in Fig~\ref{figdsored},
it is in the range $3.1-6.0\times 10^{-16}\esc$, while in the last case it only reaches $1.7\times 10^{-16}\esc$.
For a distance of $8\,{\rm kpc}$ to the Galactic center, 
the average luminosity of the observed red-shifted Br$\gamma$ line is $1.0\times 10^{-3}\lsun$, and 
twice this value when integrating over a larger aperture of radius $r=0.075''$.   

In the last example in Fig.~\ref{figdsored}
the recovered properties of the emission line are quite different to any other case, 
although the bulk of the source emission seems to
be well within the aperture placed along the orbit, and that it covers a very similar
area as the circular aperture used in the first three cases: 
e.g. the line width in the last case is narrower and  
the line flux is only $\sim 35$\% of that measured in any other background-subtracted spectrum.  
From this analysis, we call for precaution when measuring line properties along predetermined areas 
in the field-of-view.

\subsubsection{Position of the DSO}
\label{subsubsec:position}

\begin{figure*}[htb!]
\centering
\includegraphics[width=0.93\textwidth]{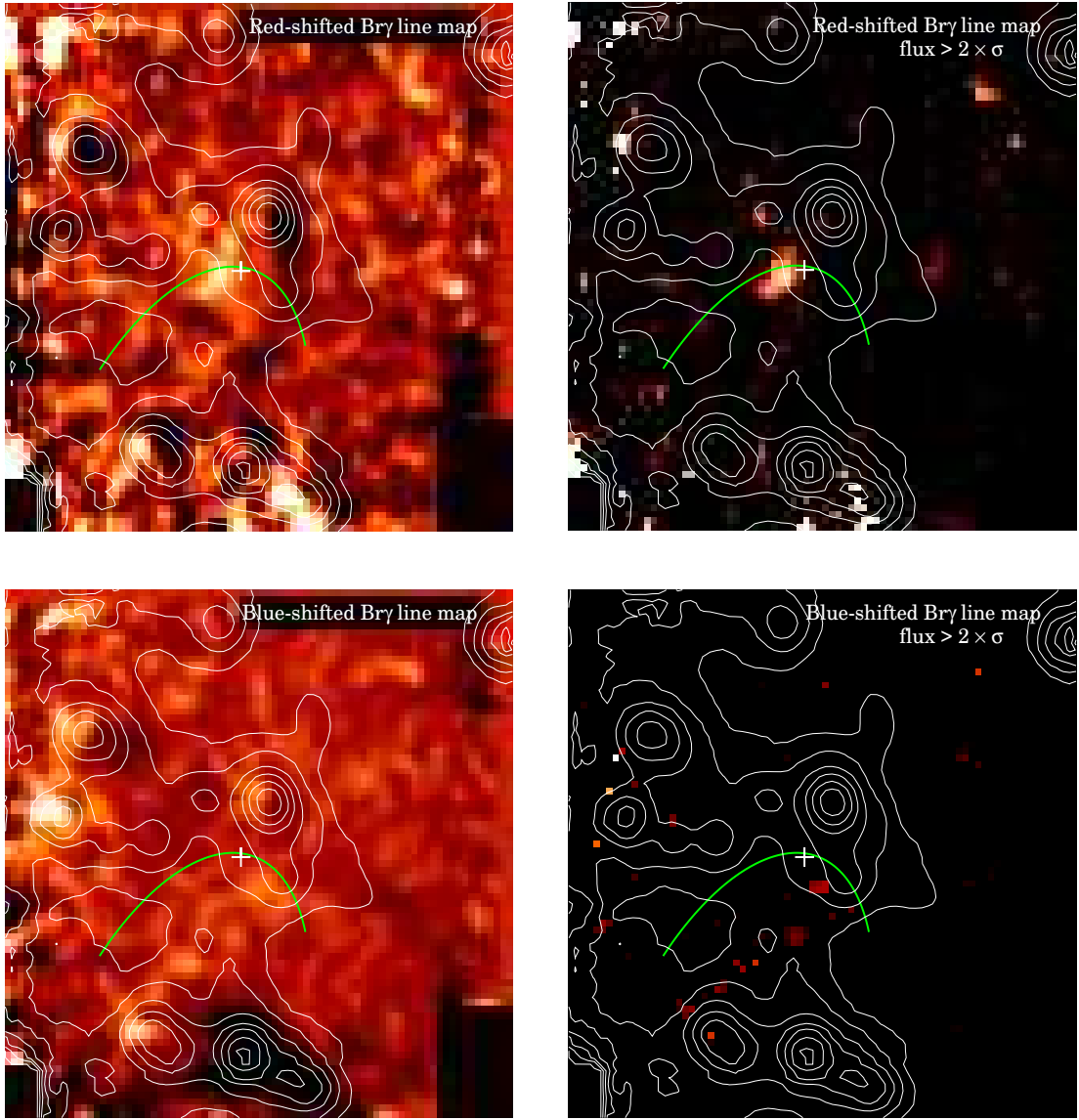}
\caption{Br$\gamma$ line maps. 
Panels show $1.0 \times 1.0\,{\rm arcsec^{2}}$ of the Galactic Center region in February-April 2014. 
The cross marks the position of SgrA*. The thick green line corresponds to the best fit 
elliptical orbit. The $K$-band continuum contours depict the brightest S-cluster members.  
{\it Top panels:}  DSO red-shifted Br$\gamma$ line map. 
{\it Left:} Integrated emission in a range of 120\,\AA  ~around $2.185\mum$ 
after subtracting the background in every spatial pixel of the field-of-view.  
{\it Right:} Same as left panel, but showing only emission that is brighter than 
$2\times$ the noise level. 
{\it Bottom panels:} Blue-shifted Br$\gamma$ line map. 
{\it Left:} Integrated emission in a range of 120\,\AA  ~around $2.147\mum$, i.e., around 
the expected blue-shifted Br$\gamma$ line emitted by a source approaching us at a speed 
of $2700\kms$.   The background has been subtracted  
in every spatial pixel of the field-of-view.  The color scale is the same as on the upper panels. 
{\it Right:} Same as left panel, but showing only emission  brighter than 
$2\times$ the noise . 
See the electronic edition of the Journal for a color version 
of this figure.
\label{figdsoredlinemap}}
\end{figure*}

To confirm the position of the Br$\gamma$ emission, we removed the  
background emission in every pixel of the field-of-view following the procedure 
described above using the spectrum shown in the second example 
of Fig.~\ref{figdsored} as a background - as classical iris photometry makes an unbiased and efficient use 
of the background in the immediate surrounding of the source.
Then we integrated the residual flux in the range $2.181 - 2.193\mum$.  
The result, shown in the top-left panel of Fig.~\ref{figdsoredlinemap}, 
is an image of the excess flux, compared to the continuum, emitted by the source in this 
wavelength range. 
Fitting a Gaussian to this emission in every spatial pixel, allows us to mask the 
areas where the flux within the line is less than $2\times$ the noise level. 
When such a mask is applied (Fig.~\ref{figdsoredlinemap}, top-right), 
the location of the DSO shown by its redshifted Br$\gamma$ emission
is clearly revealed. The position of the DSO as indicated by the position of the brightest
Br$\gamma$ peak in Fig.~\ref{figdsoredlinemap} is 8.6\,mas south and 41.5\,mas east of Sgr\,A*.

\subsubsection{The \ion{He}{1} and Pa$\alpha$ emission lines}
\label{subsubsec:otherlinered}

\begin{figure*}[htb!]
\includegraphics[width=\textwidth, bb = 20 450 580 740, clip]{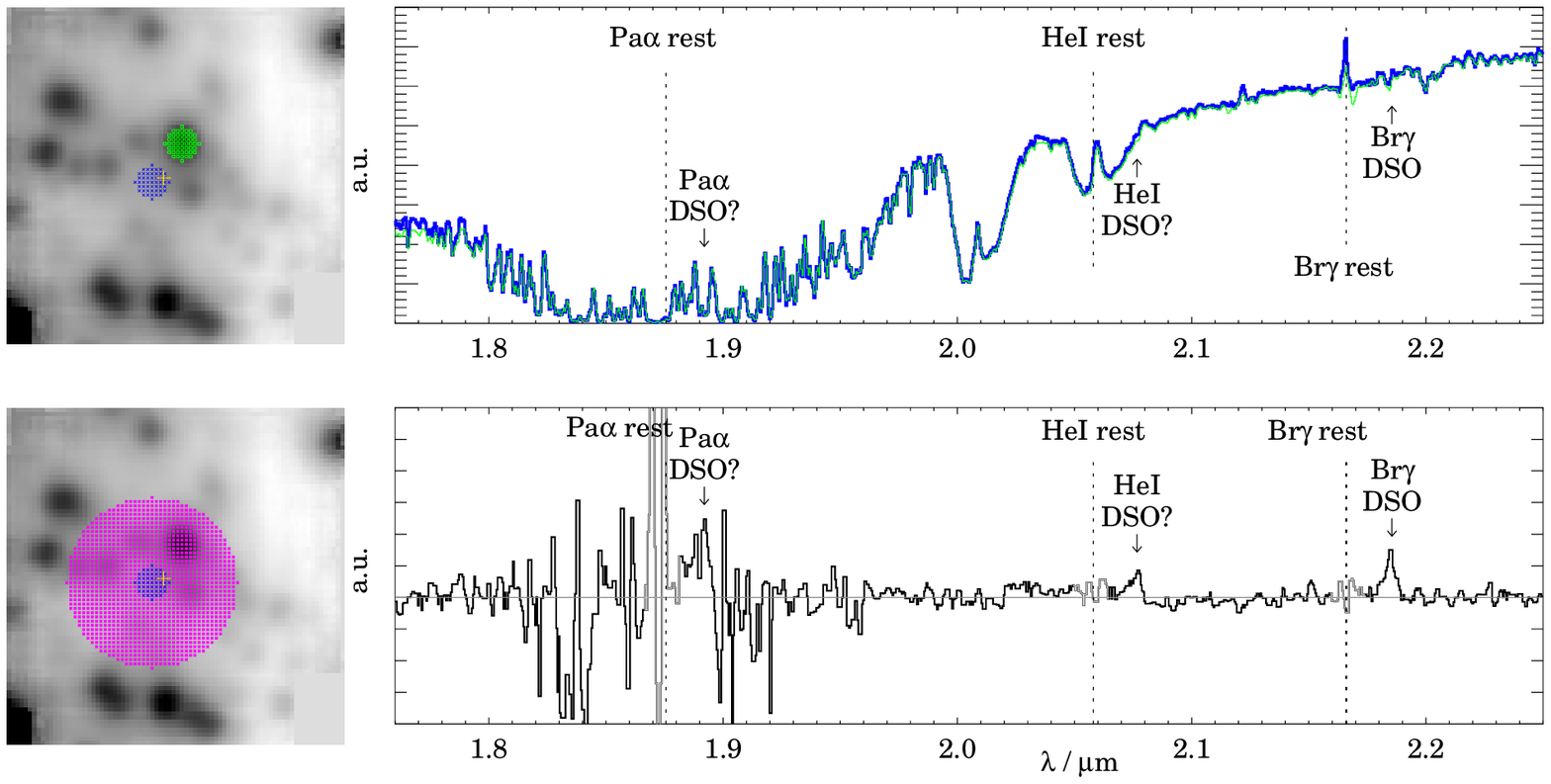}
\caption{ 
The \ion{He}{1} and Pa$\alpha$ red-shifted emission of the DSO before May.
 {\it Top left:} $1 \times 1\,{\rm arcsec^{2}}$ SINFONI mosaic of the Galactic center region in February to April 2014.
The DSO  aperture is marked with blue crosses. The spectrum extracted from the aperture placed on S2,  marked with empty green squares, 
is used to model the telluric absorption.
{\it Top right:} Comparison between the DSO spectrum (thick blue line), and the telluric model (thin green line). 
The positions of the zero velocity hydrogen and helium lines are signaled with vertical dashed 
lines, while the expected positions of the red-shifted emission lines from the DSO 
are indicated with the arrows.
{\it Bottom left:} Same as in left panels of Fig.~\ref{figdsored}.
{\it Bottom right:} Results of the subtraction of the background from the DSO spectrum, 
after correcting both spectra for tellurics.
Dashed lines and arrows are the same as above.
See the electronic edition of the Journal for a color version 
of this figure.\label{figdsoredpa} }
\end{figure*}

The detection of Pa$\alpha$ $\lambda$1.875 $\mu$m and \ion{He}{1} $\lambda 2.058\mum$ emission 
requires  modeling of the atmospheric absorption. Although we observed some 
standard stars during the different runs to use them for the telluric modeling, 
the sky variation throughout the nights was large and the corrections unsatisfactory. 
The alternative is to use a bright star in the field, or a combination of some of them, 
as tracers of the telluric absorption. 
Fig.~\ref{figdsoredpa} shows the case where the star S2 is used for this purpose. 
In the top panel, a comparison between the DSO spectrum and that of S2 is shown. 
The absorption features in the source spectrum around $1.9\mum$ are well approximated, 
 but the overall shape of both spectra differs from each other, 
as expected from the earlier discussion. 

Following the common telluric-correction procedure, the DSO spectrum 
is divided by the normalized telluric spectrum 
(in the case of Fig.~\ref{figdsoredpa} that of S2).
The same correction is applied to the background spectrum.  
Here, we selected without preference the background shown in the third example 
of Fig.\ref{figdsored}.
The resulting DSO spectrum, after the background correction, is still very noisy 
around $1.9\mum$, but  hints of the red-shifted Pa$\alpha$ and \ion{He}{1} are visible. 
The lines are observed with a signal-to-noise of about 2, in the case of helium, 
and just above one, in the case of Pa$\alpha$. 
The fact that Pa$\alpha$
is not observed with the expected strength (approximately 12 times brighter than Br$\gamma$, 
after extinction correction) is probably due to the low elevation of the Galactic 
center region in February - April which resulted in stronger telluric absorption
in this region. 
We fit the redshifted HeI with a Gaussian to obtain the line properties. 
It peaks at 2.076$\pm$0.078$\mum$, i.e. $\sim$ 2650$\pm$100 km/s within the 
uncertainties at the same receding velocity as indicated 
by the Br$\gamma$ line. It also exhibits a similar width $\sim 750\kms$. 
After correcting for extinction assuming $A_V\approx  26.8\,{\rm mag}$, 
we find \ion{He}{1} $/ {\rm Br\gamma} \sim 0.6$.  
consistent with models in which the emission is dominated 
by optically thin material \citep{gil13b,shch14}.

However, this value must be taken with caution, given the low S/N of the lines 
and the high influence of the tellurics and background corrections in the 
measured line fluxes.

\subsection{Blue-shifted pre-peribothron lines}
\label{subsec:blue}

The blue side of the line emission would be extremely diluted if we observed the source very
close to its peribothron position and the radial velocity
range would span almost 6000 km/s, significantly broadening the emission line and making 
it virtually impossible to detect. 
Only if the time span for such an event was of the order of a week (depending on the exact orbit)
observations immediately after our measurements may have picked up significant blue line emission.

\subsubsection{No detected blue Br$\gamma$ emission}
\label{subsubsec:nobluedet}

Using a variety of apertures we conducted a systematic search of the DSO blue-shifted 
Br$\gamma$ emission along the portion of the orbit which lies up-stream of the red emission. 
As we did not find blue-shifted emission at the pre-peribothron position of the DSO, 
we extended or search to consecutive positioned apertures upstream the April 2014 
pre-peribothron position.
Corresponding to the findings of 
\citet{pfu14b} the blue line emission should have been the brightest line component in our data. 
If the blue-shifted emission is compact, and at least half as bright as the red-shifted one, 
we should be able to detect it  with a $S/N \sim 2.5$ or higher. 

Following the same approach presented in Sect.\,3.1.1.,  i.e., subtracting 
background spectra created in different ways from the source spectrum, we 
aimed to find hints of the blue-shifted component.  
The DSO spectrum was integrated from an aperture with the PSF size ($r=0.05''$) 
that was placed  several times in a grid mapping a squared area of 
 200\,mas$\times$200\,mas with Sgr\,A* at the north-east corner.  

In this way we covered the large area to the west
and to the south of SgrA* where the approaching side of
the DSO is expected to be found.
We also searched for the blue-shifted emission using slightly larger apertures to account for a possibly  
more extended emission that could be expected in case the source 
was not as compact as before the peribothron passage. 

\begin{figure*}[p!]
\includegraphics[width=\textwidth, bb = 20 160 580 740, clip]{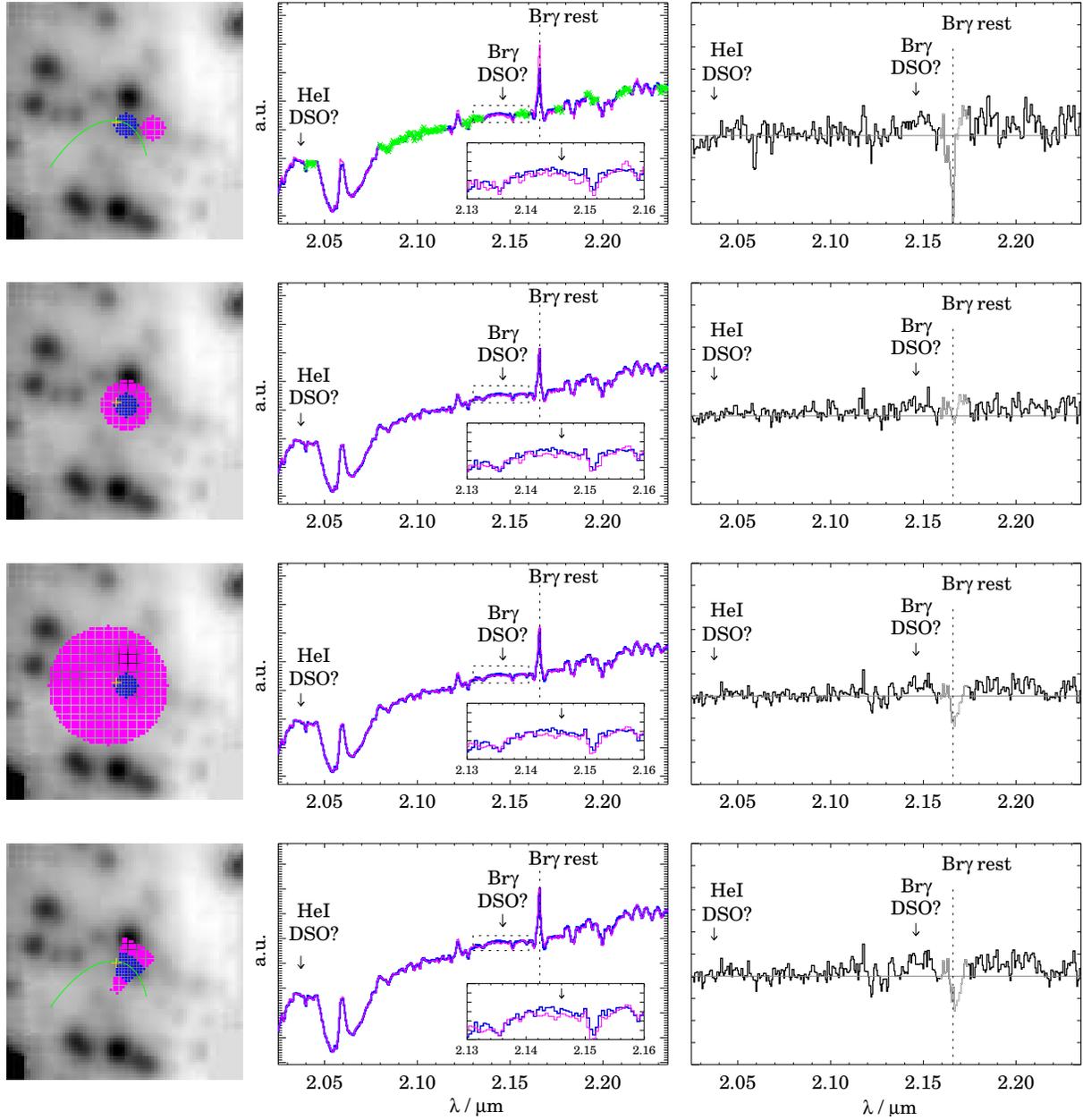}
\caption{Non-detection of Br$\gamma$ blue-shifted emission before May at a first up-stream position
at 43\,mas east and about 11\,mas south of SgrA*.
{\it Left panels:} Same as Fig.~\ref{figdsored}.  
{\it Central panels:} Comparison between the DSO (thick blue line) and the background (thin pink line) spectra.  
Arrows at $2.037\mum$ and $2.146\mum$ indicate the approximate location of the expected 
DSO blue-shifted \ion{He}{1} and Br$\gamma$ emission lines, in case the speed 
of the approaching component was the same as the receding one. 
The inset panel corresponds to the dashed-line box, which is a zoom-in to the spectra in the $2.13-2.16\mum$ range. 
The arrow in the inset panel marks again the position of the blue-shifted Br$\gamma$ line.  
The spectral windows marked with crosses in the top panel are used to fit the slope of the background spectrum
to that of the DSO. 
{\it Right panels:} Results of the subtraction of the background from the DSO spectrum.
The vertical range of the plots corresponds to one unit in the middle panels 
and spans the same range as in the right panels of Fig.~\ref{figdsored}.
The zero-line is shown in gray. 
See the electronic edition of the Journal for a color version 
of this figure. \label{figdsoblue1}}
\end{figure*}

\begin{figure*}[p!]
\includegraphics[width=\textwidth, bb = 20 160 580 740, clip]{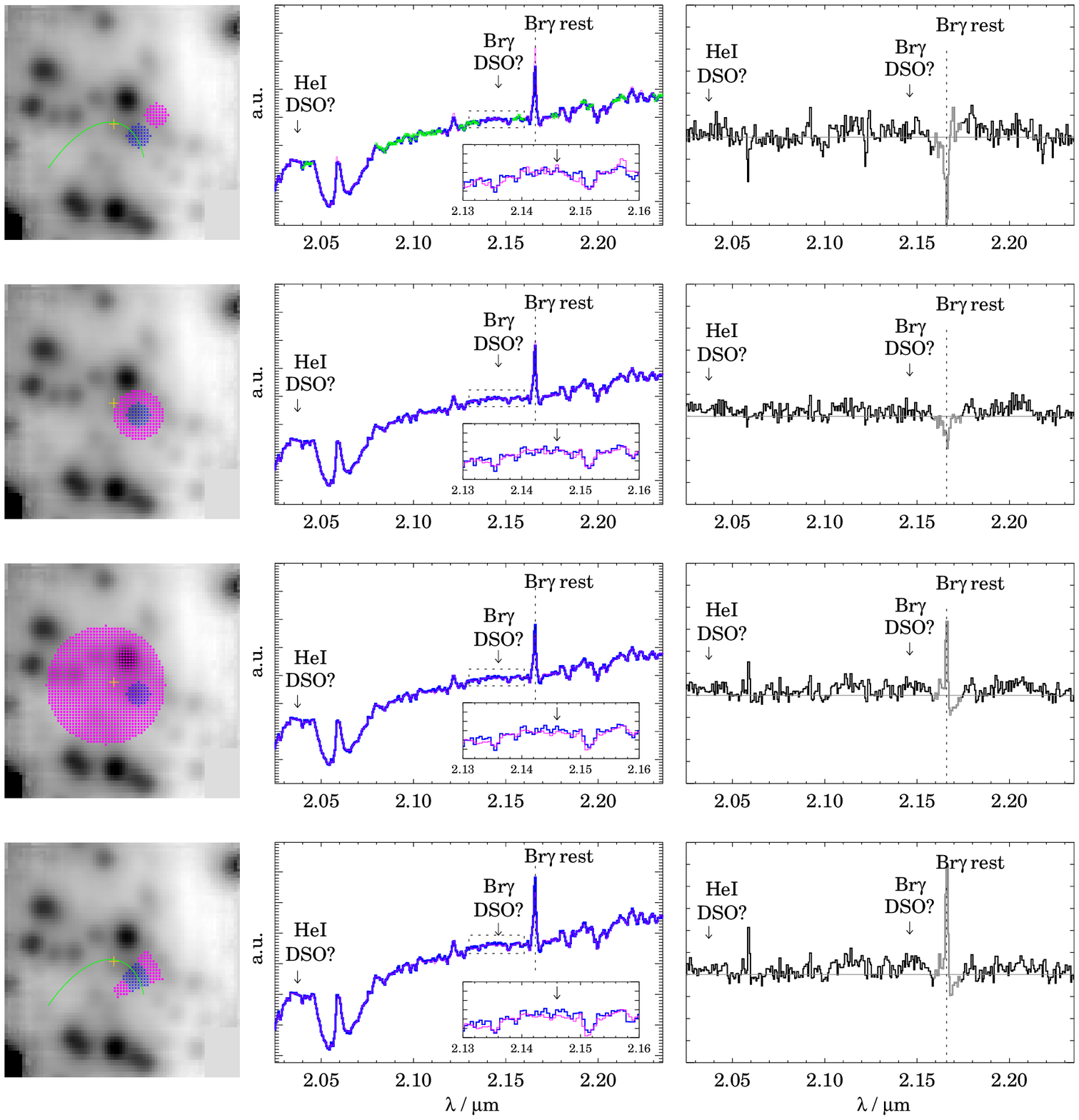}
\caption{Non-detection of Br$\gamma$ blue-shifted emission before May at a second up-stream position
at 103.8\,mas West and 50\,mas South of SgrA*.
Panels and symbols are the same as in Fig.\, \ref{figdsoblue1}.
See the electronic edition of the Journal for a color version 
of this figure. \label{figdsoblue2}}
\end{figure*}

Figs.~\ref{figdsoblue1} and \ref{figdsoblue2} show two attempts of 
finding the Br$\gamma$ DSO emission 
in two different positions upstream of the best fit elliptical orbit.
These are examples of the systematic rearch for the DSO Br$\gamma$ emission
south/west of SgrA*. The apertures are placed at the position
(and one consecutive position) along the orbit 
at which \citet{pfu14b} and earlier \citet{gil13b} had reported the detection of blue-shifted line emission.

As in Fig.~\ref{figdsored}, left panels show the size and position of 
the background and source apertures, middle panels compare the 
spectra extracted from them, and right panels present the 
subtraction of the two.  
The expected spectral positions of the blue-shifted Br$\gamma$ and \ion{He}{1} are 
derived assuming the emitting source to approach us after peribothron 
with a similar speed as the still receding part. 
The vertical range in the right panels is the same as the one used in Fig.\ref{figdsored}, 
and therefore can be directly compared. 
In case there is a source emitting a blue-shifted line at any of these two positions, 
the line should be clearly visible in all four rows displayed within one figure. 
This is because, in these four examples, the source spectrum is extracted from the 
same region and the only difference must be the signal to noise
ratio of the line which depends on the subtracted background 
(we come back to this point in section \ref{subsubsec:background}).
Hence we can rule out that there is a blue
line similar to that seen on the red shifted side de-
spite the fact that the source should be similarly
compact (see Figs. 1 and 5 in \citet{pfu14b}
and comments in our introduction section).

\subsubsection{ Upper limit for the blue-shifted Br$\gamma$ line}
\label{subsubsec:upperlimit}

Another strategy for searching a line emission is, as we did in previous sections, 
to subtract a background spectrum from the entire data cube and then
integrating the remaining flux within narrow spectral windows around the 
expected wavelength.   
For this search we used different background spectra and integrated the 
residual flux in the range $2.143 - 2.151\mum$. 
Fig.~\ref{figdsoredlinemap} (bottom-left) shows one example.  
We fitted a Gaussian to every spatial pixel to create a mask that 
selects those places where the flux is less than $2 \times$ the 
noise level. In the right panel of Fig.~\ref{figdsoredlinemap} (bottom-right), such 
mask has been applied.  
We see possible hints of a spatially compact source at 37.5\,mas west and 68.8\,mas south 
of SgrA* that is not located on the expected DSO orbit. 
Looking at the line properties we find that in average the emission is very broad, 
with FWHM$>2\,000\kms$ (i.e., larger than $0.015\mum$), 
and centroid at $\sim 2.149\mum$.  
Assuming that the blue-shifted Br$\gamma$ line emission is as wide as the 
red-shifted one, i.e. $720\,\kms$, and with a noise in that spectral range 
of $\sim 2.9\times 10^{-14}\,{\rm erg\,s^{-1}\,cm^{-2}\,\mu m^{-1}}$, we 
obtain an upper limit for the line flux of $\sim 4.7\times 10^{-16}\esc$, i.e.
a luminosity $L({\rm Br\gamma_{blue}}) < 1.0 \times 10^{-3}\lsun$. 
Whether this emission is real considering the multiple sources of noise, and 
whether it has some relation with the DSO is unknown. 
For the apertures placed along the orbit, the upper limit of  a blue-shifted 
Br$\gamma$ line-flux is  $\sim 2.8\times 10^{-16}\esc$, which is about a half 
of that of the red-shifted line.

\subsubsection{Influence of the selected background }
\label{subsubsec:background}

\begin{figure*}[p!]
\includegraphics[width=\textwidth, bb = 10 420 580 740, clip]{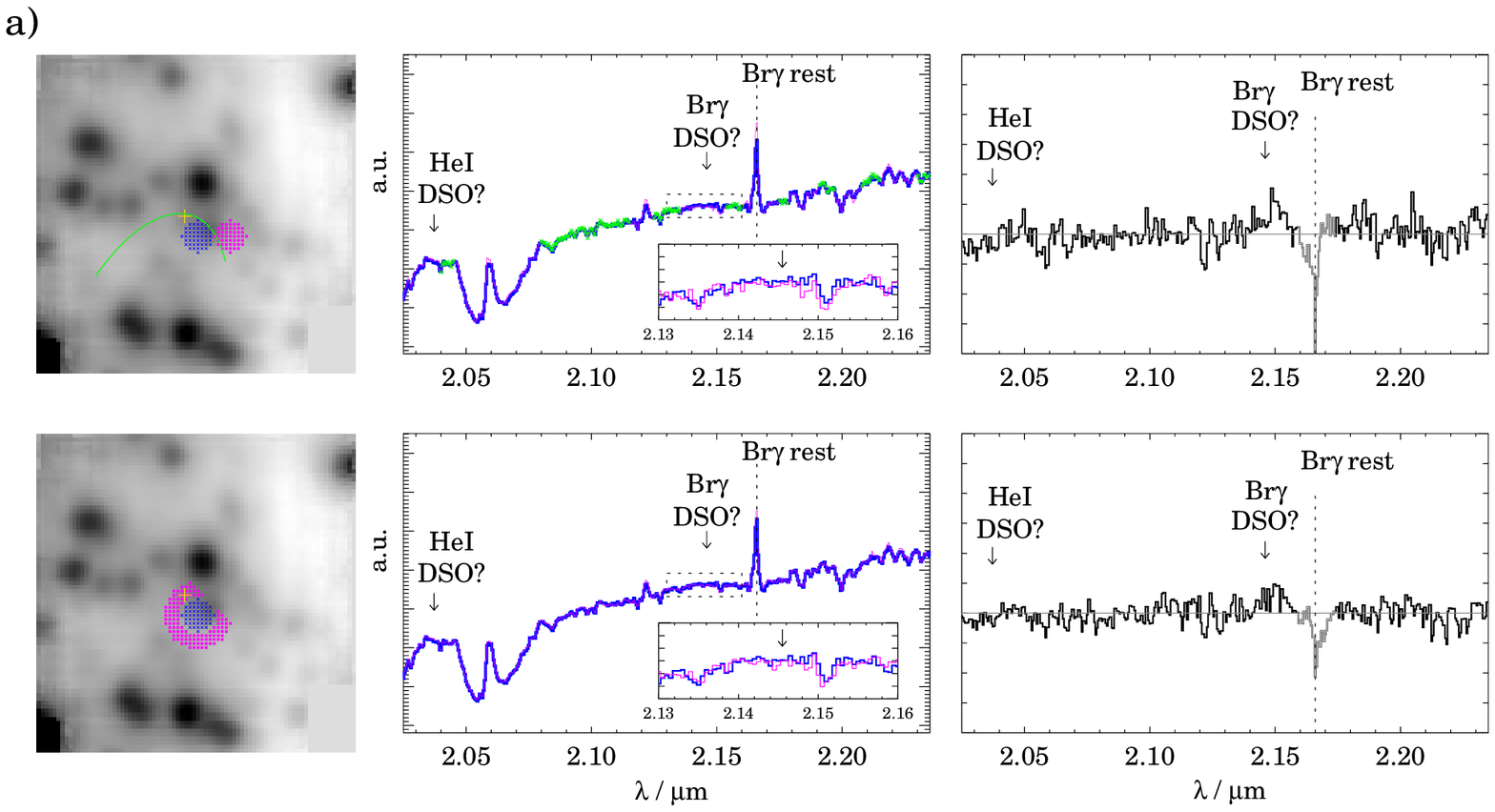}
\includegraphics[width=\textwidth, bb = 10 450 580 740, clip]{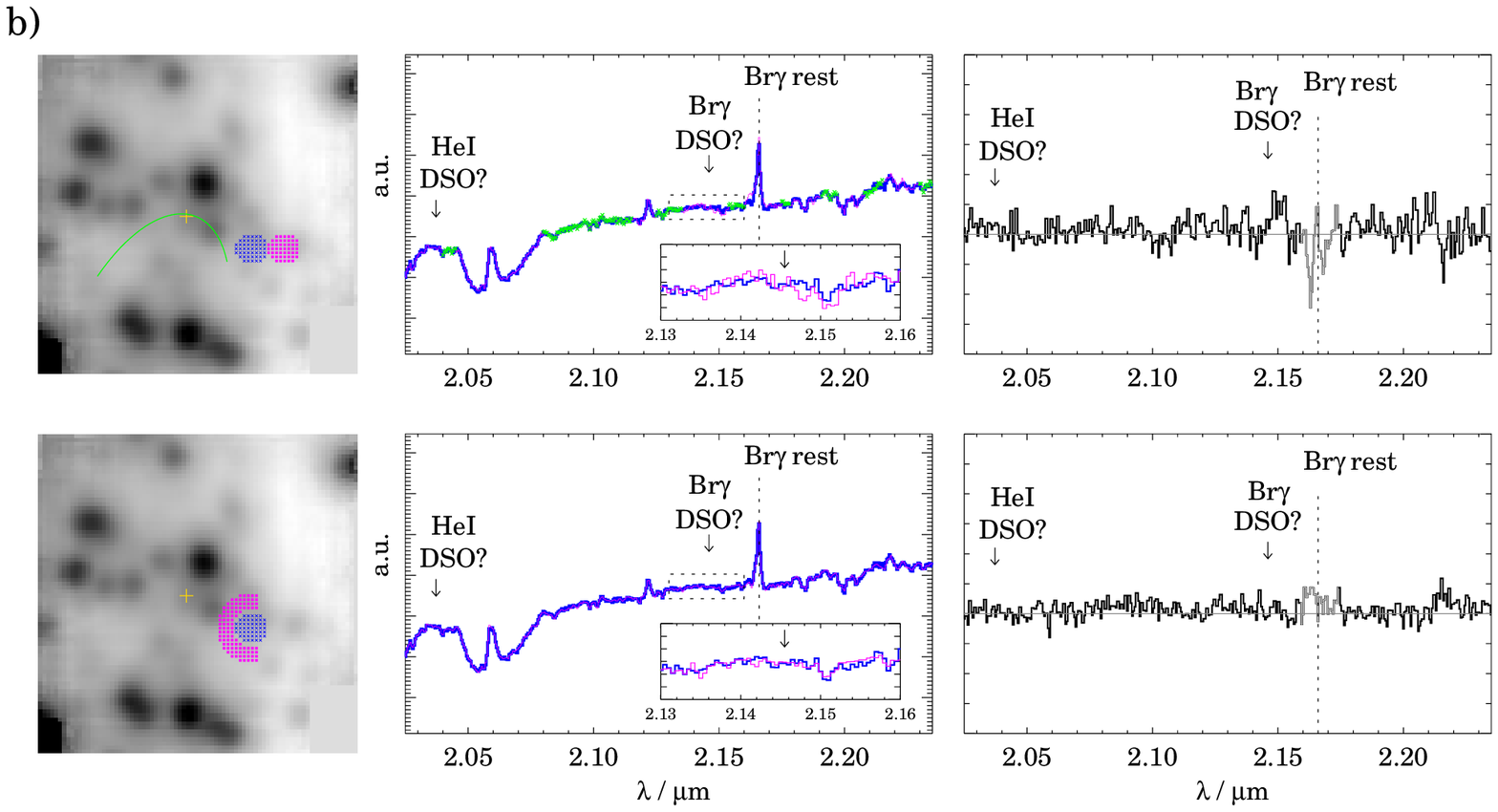}
\caption{ Spurious blue-shifted Br$\gamma$ emission due to background selection. Panel descriptions are 
the same as in Fig.\,\ref{figdsoblue1}. The two shown cases correspond to sources at 
{\it a)}37.5/,mas west and 68.8/,mas south of SgrA*, and {\it b)} ~200\,mas west and 100\,mas south of SgrA*.
See the electronic edition of the Journal for a color version of this figure. 
\label{algo}}
\end{figure*}

There is no doubt that the subtraction of the background emission plays 
an important role in the detection of faint line emission. 
The usage of different background spectra from regions close to the 
position of interest is an effective tool to discriminate between a source line emission 
and the unlucky presence of a background feature at the studied wavelength.
In the first row of Fig.\,\ref{algo}a and \,\ref{algo}b we present examples of background that produce 
a spurious blue-shifted Br$\gamma$ emission at positions far away from the expected orbit.  
The panels in the second row of each example show how, after selecting different 
background spectra, a very good overlap with the source spectrum is obtained 
and only noise remains after the subtraction.
The aperture shown in Fig.~\ref{algo}a is located at the position
of the bright blob closest to SgrA* in the bottom panels
of Fig.~\ref{figdsoredlinemap}.
The background subtracted spectrum of this aperture
was used above to estimate the upper limit for the blue-shifted line.
In this case, the S/N of the feature at $\sim 2.15\mum$ depends strongly
not only on the background selected, but also on the way it is
scaled and subtracted.
As we were not able to produce spurious detections on the red side we conclude that 
those in the blue may result from an enhanced local variation of the background 
in this particular spectral range.
Based on this analysis, we call for caution  when studying line emission properties of 
faint sources in crowded fields.

\begin{figure*}[htp!]
\includegraphics[width=\textwidth]{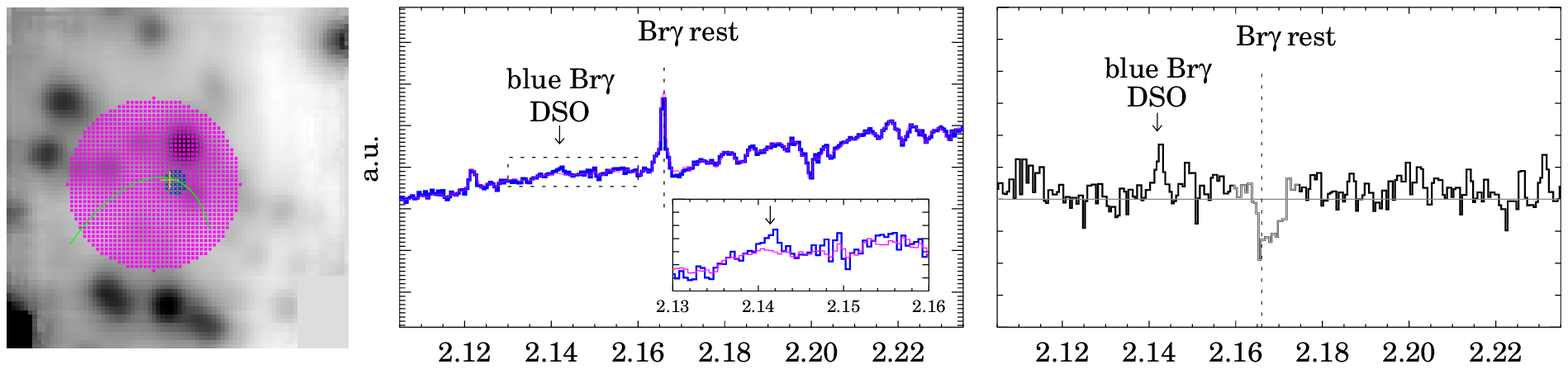}
\includegraphics[width=\textwidth]{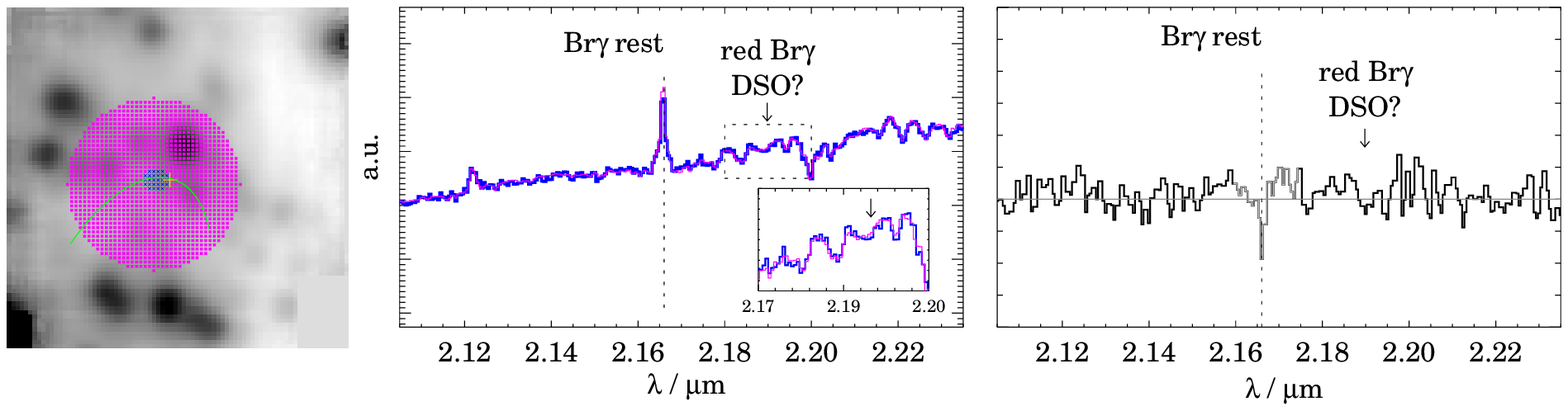}

\caption{Br$\gamma$ line emission after peribothron passage: 
Top: Blue-shifted Br$\gamma$ emission detected
at 12\,mas west and about 5~mas north of SgrA*.
Bottom: No red- or blue-shifted Br$\gamma$ emission detected at our pre-peribothron position
at 43\,mas east and about 5\,mas south of SgrA* .
See the electronic edition of the Journal for a color version of this figure. 
\label{figdsoblueafter}}
\end{figure*}

\begin{figure*}[hbt!]
\centering
\includegraphics[width=\textwidth]{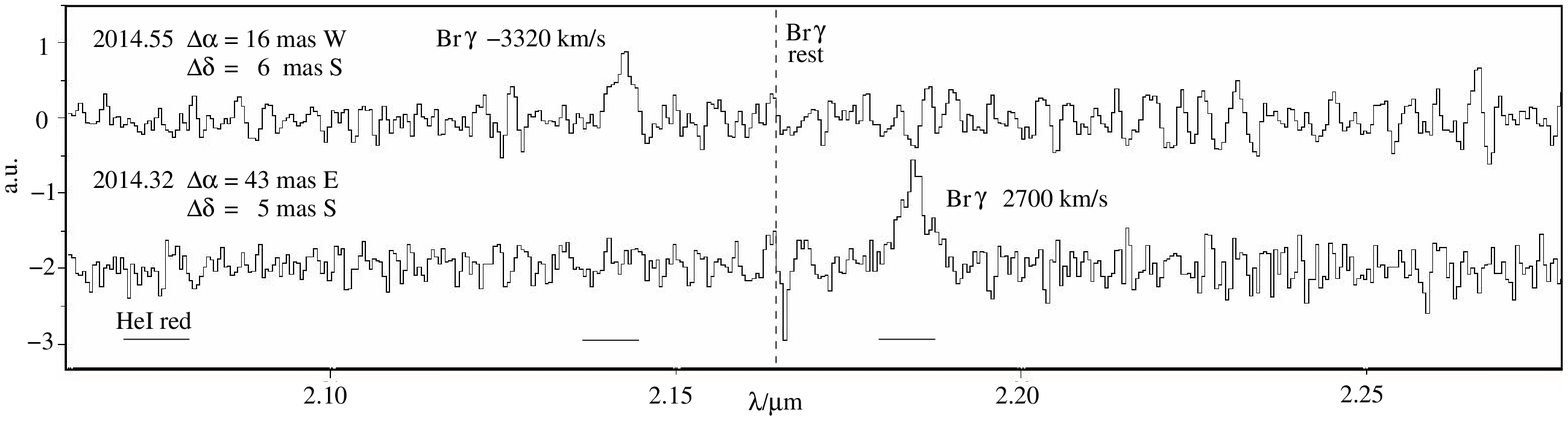}
\caption{Summary spectra at the pre- and post-peribothron positions for the April and June epoch 2014.
See details in text.
\label{fig:summaryspectra}}
\end{figure*}

\subsection{Blue-shifted post-peribothron Br$\gamma$}
\label{subsec:redpost}
Fig.~\ref{figdsoblueafter} shows the spectrum integrated over an aperture of radius $0.05''$
at the post-peribothron in June 2014 at position at 16\,mas west and about 6\,mas south of SgrA*
at a S/N of 2.5 to 3.1 depending on the background subtraction.
The line has a blue-shifted center velocity of -3320$\pm$60~km/s and after correcting for 
spectral resolution a FWHM of 15$\pm$10~\AA. 
Line flux and width were derived using several background corrections similar to
what is shown in Fig.\ref{figdsored}.
The Br$\gamma$ line luminosity is about $0.4\times 10^{-3}\lsun$.
The narrow line estimate could be a result of the weak line detection, it could also point at
a stellar nature of the source (see below).

The excess line emission can clearly be seen even before background subtraction in the inset of the middle top plot 
in Fig.~\ref{figdsoblueafter}.
Based on the post-peribothron observing dates listed in Tab.~\ref{tab1}
we assigned an integration time weights epoch of 2014.55 to this measurement.
No red-line emission can be claimed for this epoch at this position.
In the lower plots of Fig.~\ref{figdsoblueafter} we show that  at that epoch neither red- nor blue-shifted 
Br$\gamma$ line emission can be see at our pre-peribothron position (see also the inset of the middle bottom plot).
The excess line emission can clearly be seen even before background subtraction in the inset of the middle top plot.
The baseline used for this spectrum excludes the region around blue-shifted 
(2.138-2.146~$\mu$m) and the red-shifted Br$\gamma$ and He line emission
(2.175-2.190~$\mu$m and 2.070-2.080~$\mu$m)
we used at the pre-peribothron position.
No red-shifted emission was detected at the post-position.
No red-shifted line emission was detected at any position down-stream of the post-peribothron position.
In Fig.~\ref{fig:summaryspectra} we show summary spectra at the pre- and post-peribothron positions for 2014.
We obtained the spectra using a 0.050'' radius source and a 0.25'' radius background aperture 
centered on the DSO. We subtracted a high-pass filtered version of the spectra that we obtained 
by replacing the range over which detectable line emission is present 
(indicated by the three lines at the bottom of the graph) by the mean in the neighboring spectral elements 
and smoothed the resulting spectrum with Gaussian having a FWHM of 10 spectral resolution elements.
The location of the Br$\gamma$ rest emission is indicated by a vertical dashed line.

\begin{figure}[hbt!]
\includegraphics[width=0.99\linewidth]{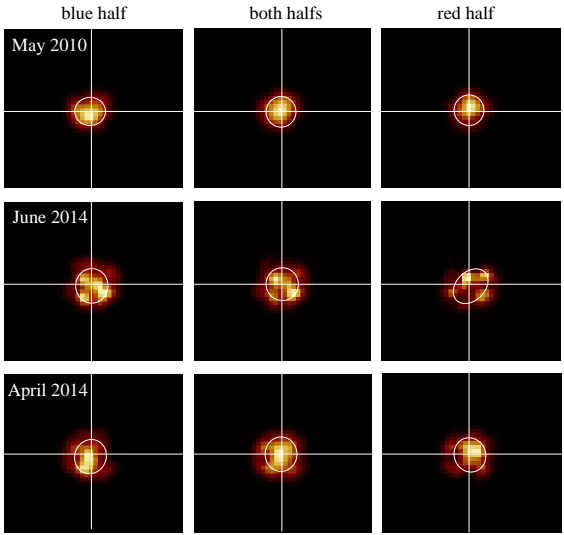}
\caption{Maps of the DSO in its Br$\gamma$ line emission for the 
April 2014.32,
June 2014.55 and
May 2010.45 
epochs.
The maps are 0.5''$\times$0.5'' in size.
The centroid data are given in Tab.~\ref{tab:linecentroids}.
See the electronic edition of the Journal for a color version of this figure. 
\label{fig:linecentroids}}
\end{figure}

An important question is that of the size of the line emission region and possible velocity gradients across the DSO.
To investigate this we obtained line maps of the Br$\gamma$ emission.
In Fig.~\ref{fig:linecentroids} we show maps of the DSO in its Br$\gamma$ line emission 
for the times before (epoch May 2010.45\footnote{SINFONI data from ESO program 183.B-100} and April 2014.32) and after (epoch 2014.55) the peribothron.
For the brightest and least confused Br$\gamma$ line maps for May 2010 (Fig.~\ref{fig:linecentroids})
we find a geometrical mean FWHM of 6.5~pixels. 
For the star S2 we find a FWHM of 6.2~pixels. With 12.5~mas per pixel this gives an upper limit 
on the deconvolved FWHM source size of 24~mas. 
The centroid positions of the emission line maps of the left half, right half and full line 
in milliarcseconds relative to the position of the full line map centroid position are given in Tab.~\ref{tab:linecentroids}.
Under the assumption that differences in the relative positions of the red- and blue-half of the 
single line Br$\gamma$ emission line map can be interpreted as being due to a velocity gradient
of a tidally streched source we find for all epochs an upper limit of the corresponding source size of 15~mas.
This implies that the source emitting the bulk of the Br$\gamma$ line is very compact
and we adopt the value of 15~mas as an upper limit on the line emitting FWHM source size.
This is consistent with the analysis of L'-band continuum images by  \citet{eck13} 
showing that $>$90\% of the DSO emission at 3.8$\mum$ wavelength is compact (FWHM $\le$20~mas) 
and only up to 10\% of the flux density of the DSO can be extended on the scale size of the PSF. 
Our size limit is also consistent with the upper limit of 32~mas presented by \citet{witzel14}.
These size estimates are all smaller or at the lower bound of the 2008-2013 size estimate 
of 42$\pm$10~mas \citet{gil13b}.
Our adopted Br$\gamma$ source size corresponds to 120~AU at a distance of 8~kpc, i.e. it is of the
close to the peribothron distance of the source.
However, it is still about 50 times larger than the estimated size of an optically thick dust shell
of a 2\solm ~star of about 2.6~AU \citet{witzel14}.

\section{Discussion}
\label{sec:Discussion}

The fact that we found red-shifted Br$\gamma$ emission at the pre-peribothron position
but did not detect any blue-shifted emission up-stream
(in sections \ref{subsubsec:upperlimit} and \ref{subsubsec:nobluedet}
we only show two examples - we probed several positions along the orbit) 
and vice versa (section \ref{subsec:redpost}) has implications on the orbit and on the DSO model.

\begin{deluxetable}{rrrrrrrrrr}
\tabletypesize{\scriptsize}
\tablecaption{Centroids of DSO Line Maps \label{tab:linecentroids}}
\tablewidth{0pt}
\tablehead{
\colhead{epoch} & \multicolumn{2}{c}{blue half}&  \multicolumn{2}{c}{full line} &   \multicolumn{2}{c}{red half} &   \multicolumn{2}{c}{red-blue half}
\\
          &
[mas]&
[mas]&
[mas]&
[mas]&
[mas]&
[mas]&
[mas]&
[mas]
\\
}
\startdata
April 2014&$-$6.75&$-$7.63 & 0.0   & 0.0  &+6.25 & +6.00 & +13.0   &+13.6\\
June  2014&$-$0.25&  +4.25 & 0.0   & 0.0  &+4.50 &$-$3.88&$-$4.75  & $-$8.13\\ 
May   2010&$-$4.63&$-$6.00 & 0.0   & 0.0  &+4.13 & +4.63 &  +8.75  &+10.63 \\ 

\enddata
\tablecomments{~
For the single emission line we measured at both 2014 and the May 2010 epochs we list the line map right ascension 
and declination centroids for the blue and red half 
of the corresponding single line in milliarcseconds  
with respect to the 'full line' position. 
In the last two columns we list the positional difference between the red- and blue-half centroid positions. 
One pixel corresponds to 12.5~mas.
}
\end{deluxetable}

\begin{figure}[hbt!]
\centering
\includegraphics[width=1.1\linewidth]{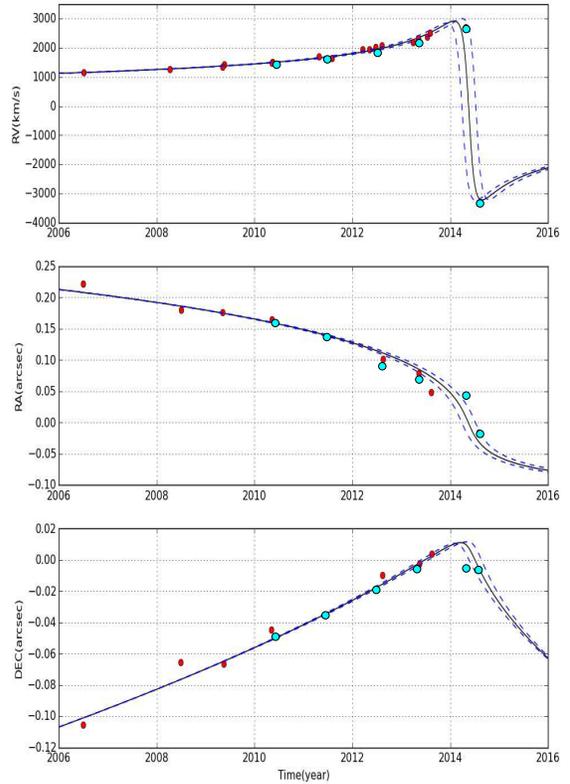}
\caption{Right ascension, declination and radial velocity 
of the DSO together with the best orbital fit we obtained.
See details in text.
\label{fig1}}
\end{figure}

\begin{figure}[hbt!]
\centering
\includegraphics[width=1.1\linewidth]{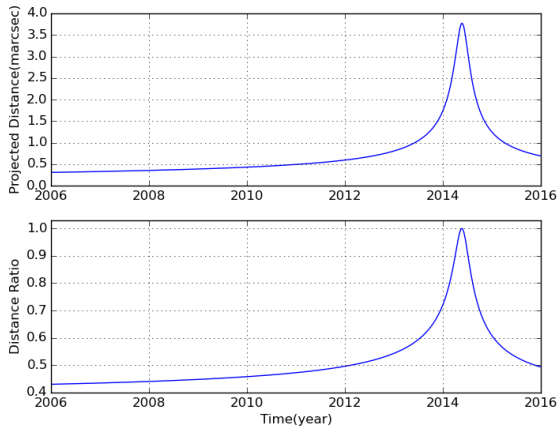}
\caption{Orbital projection effects:
Top: The evolution of the projected separation between two neighboring points 
of arbitrary 0.5 units in 2011.
Bottom: Foreshortening factor of any structure along the orbital extent as a function of time.
\label{fig:fore}}
\end{figure}

\subsection{The orbit}
\label{subsec:orbit}

Based on L-band imaging an infrared excess source within the central cluster of high velocity S-stars was
found to approach the immediate vicinity of SgrA* \citep{gil12}.
In addition, Br$\gamma$ line emission was reported by \citet{gil13a} and \citet{phi13}. 
In \citet{eck13} we report the identification of K-band emission from a source at the position 
of the L-band identifications. 
\citet{gil13b} report a marginally spatial extension of the Br$\gamma$ line emission in their SINFONI data
and find an intrinsic Gaussian FWHM size of $42\pm10$~mas (using 2008-2013 data). 
Given the peculiar orientation of the source estimated 
orbit, precise estimates of the source elongation along the orbit are difficult to obtain.
Combining these observational facts indicated that a dusty object - possibly associated with a stellar object -
is on an elliptical orbit around SgrA*.
The observational data were also used to derive the orbit of this object and to predict its peribothron transition.
Due to the presumably high ellipticity of the orbit only very weakly curved sections of the orbit were available 
and first predictions of the peribothron transition time in 2013 \citep{gil12} proved to be incorrect.
The inclusion of (or even restrictions to) the Br$\gamma$ line emission resulted in new predictions for
early 2014 \citep{gil13a, phi13}. 
The fact that the telescope point spread function (PSF) in the L-band is intrinsically larger and therefore
more susceptible to diffuse extended emission is probably the main reason for this discrepancy.

However, the predicted interactions of the gas and dust with the strong gravitational field of SgrA*
have shown that the gas itself may also not be a good probe of the exact orbital motion.
This is supported by the spatial extent and the velocity gradient across the Br$\gamma$ line emission.
It is also highlighted by the expected interaction of the DSO with the ambient medium and the gravitational field.
Therefore, even though the recently derived Br$\gamma$ based orbital solutions are in reasonable agreement
\citep{mey14a, mey14c}, the orbital elements may still be uncertain. 

Using results from our measurements with SINFONI obtained between February and September 2014, 
SINFONI archive data as well as the published Keck data \citep{mey14a, mey14c} we revisited 
the determination of the DSO orbit.  
Given that the red emission is only about 40~mas East of SgrA* and at a radial velocity of 
about 2700$\pm$60~km/s and blue emission about 30~mas West of SgrA* at $-$3320~km/s 
we obtained a new orbital solution which places the peribothron passage at 2014.39$\pm$0.14.
a bit later but close to 2014.2 as derived earlier \citep{mey14a, mey14c}.
Otherwise, the orbital elements are very similar to the ones derived earlier. 
In Fig.\ref{fig1} we show the fit to the data we used. 
The formal statistical uncertainties of the positional measurements are of the order of a few milliarcseconds.
However, the systematic effects probably limit the uncertainties to a value closer to $\pm$10~mas.
(see Fig.9 by \citet{eck12a}).
For the 2014 data presented here the exact positioning of the Br$\gamma$ line emission
critically depends on the transfer of the SgrA* position as obtained during a 3~mJy flare we observed
in March to all the other 2014 epochs. This was done using the known position and velocity of the 
southwards moving bright star S2 that is currently about 0.11'' north and 0.06'' west of SgrA*.
For the radial velocities we assume the value of 60m/s we adopted for the SINFONI data.
The light blue filled circles indicate our per- and post-peribothron 2014 data points 
we obtained using SINFONI at the VLT.
The other light blue filled circles show the results from our re-reduction of earlier SINFONI VLT archive data.
Red filled circles represent data as obtained with the Keck telescope and published by \citet{mey14a,mey14c}. 
The dashed lines indicate the approximate 1$\sigma$ 
uncertainty of the fit. 
The orbital elements are given in Tab.~\ref{elements}.
With the ellipticity $e$=0.976 and the half axis 
length of 33~mpc we obtain a peri-center distance of about 163~AU.
Which is comparable to  previous estimates \citep{pfu14b, phi13, mey14a} 
and indicates that even if the DSO is an embedded star, its outer shell may very well be 
subject to tidal disruption \citep[see also section \ref{sec:origin} and ][]{eck13, witzel14}.

In Fig.~\ref{fig:fore} we show the size evolution of structures along the orbit under the
assumption of freely moving neighboring points.
In the case of the DSO the shapes of 
the two graphs by chance look very similar. We verified that they are indeed very different 
for other orbital configurations, i.e. lower inclination or the apobothron pointing towards the observer.
The top graph shows the evolution of the projected size of a source moving along the orbit.
The bottom graph shows the same quantity divided by the actual three dimensional size of the
source, i.e. the amount of foreshortening that the observer needs to correct for.
Both graphs demonstrate that close in time to the peribothron passage the foreshortening correcting is
close to unity and that the DSO can be seen close to its full extent along the orbit.

\subsection{Tidal interaction with SgrA*}
\label{subsec:gravinter}

The way in which the gas cloud will get disrupted depends on the exact orbit and the 
nature of the DSO, i.e., if there is a stellar core or not. 
If there is a central star, then higher mass (typically ten solar masses) objects may 
retain more of the gas and dust mass in their corresponding Roche lobe than low mass 
objects (one solar mass and below). 
This is discussed in \citet{eck13}.
Recent model calculations for cases with a stellar core or even a binary core have 
been published by \citet{zajacek14}

A tail that is physically connected to the DSO has been reported by \citeauthor{gil12} (\citeyear{gil12, gil13a, gil13b}).
\citeauthor{eck13} (\citeyear{eck13, eck14}), \citet{phi13}, and \citet{mey14a} have questioned this physical association 
of the DSO with the extended Br$\gamma$ and dust continuum emitting filament about 0.3'' SEE of SgrA*.
The rather extended shape of this emission close to rest frame velocities 
may very well be associated with the Galactic center fore/back-ground features which
are numerous in this region.
It also does not follow precisely the orbital track of the DSO.
Especially at velocities close to rest frame velocities the general central cluster region is very crowded.
Hence, despite an indication of very faint emission pointing towards this general region 0.3'' SEE of SgrA*
a physical association of the bright 'tail' emission is still questionable.

If the DSO is a pure very compact and solitary gas and dust cloud, then it formed through a special 
process at a very special place and time. 
As speculated by \citet{pfu14b} it must have formed between 1990 and 2000. 
Unless one claims, as a further special feature of this source, that it has been formed at a 100\% efficiency, some 
relics and further similarly compact dust filaments or bullets must have been formed along that process. 
These have not been identified yet. It also must be noted that during the 1990-2000 time interval
the entire Galactic center region 
was under detailed investigation in the entire NIR/MIR and radio wavelength band with observing runs closely placed in time.
No special event in the mini-spiral to the south-east of SgrA* had been reported then.

It has been noted that the thermal instability can explain in a natural way the 
pressure equilibrium between the hot and the cold plasma in the mini-spiral region 
\citep{czerny2013, rozanska14}.

Hence, this process is relevant for the possibility of survival of the infalling clouds in the region, and it also allows us to estimate the typical size of clouds. In fact, clouds as large as $10^{14}$-$10^{15}$cm (0.001-0.01 light years) 
can persist. 
From the dominating optically thin DSO line emission a total mass of the clouds of $\simeq10$ Earth masses can
be derived, depending on strength of the ambient radiation field. This agrees with recent results \citep{shch14}
suggesting the mass of the DSO/G2 cloud to be within the range $4$--$20M_\oplus$. Naturally, this would be upon the assumption of a core-less cloud scenario, whereas the mass estimates do not apply if a star is embedded within the cloud. It turns out that clouds located in the distance exceeding $\sim0.05$ pc from Sgr A* can survive a few hundred years, which means that the cooling and evaporation time is shorter than the free-fall time onto the black hole.

\citet{gil13b} and \citet{pfu14b} suggest that 13 years prior to the DSO peribothron passage the source G1 went through 
its   on an orbit connected to the current DSO orbit. 
Dust heating of G1 would then explain the moderate infrared excess of 
0.3 K-band magnitudes of the star S2 in 2002 \citep[appendix by ][]{pfu14b}
as it passed close to SgrA* and G1.
However, with \citet{sabha12} we have shown that over timescales of a few months to a few years - especially close to the center -
serendipitous sources can frequently be formed due to density fluctuations of the background stars in the central 
arcsecond.

The DSO is supposed to be on a similar orbit as the G1 source \citep{pfu14b}.
However, up to this point, as it is approaching peribothron ,
the DSO has not shown any increase in K- or even L-band flux density that could be attributed to dust heating.
In fact, while approaching SgrA* in 2013, the L-band identification of the DSO has been lost \citep{pfu14b}.
This can in part be attributed to confusion. 
To some extent diffusion or destruction of dust as the source entered the immediate surroundings of SgrA* may be 
responsible as well, but certainly no brightening of the source in its K- or L-band emission has been observed.

Until now several models place a star at the center of the DSO \citep{mur12, eck13, sco13, bal13, phi13, zajacek14}.
An at least partial tidal disruption is also expected if the DSO is an embedded star.
A Roche description of the SgrA*/DSO system \citep{eck13} suggests that a more massive central
star will lose less of the gas and dust from the central few AU than a solar mass type star or a dwarf.
Simulations of compact systems by \citet{zajacek14} support this finding as well
(see also section \ref{subsec:ttau}).

For source sizes that are much smaller than the peribothron distance with \citet{jalali14} 
we have shown that at the peribothron position the gaseous source volume is
compressed by at least a factor 2 due to gravitational focusing.
This results in the fact that before and after peribothron the source stays relatively 
compact despite the influence of possible turbulences and shocks that may be induced due to 
shearing gas streams close to peribothron .
Depending on the density of the overall environment hydrodynamic interactions with 
the ambient material set in well past peribothron .
This is consistent with all hydrodynamic and particle simulations that have been used to predict the 
future development of the DSO or similar sources
\citep[e.g.][]{zajacek14, bur12, sch12, jalali14}.

\begin{deluxetable}{cccccccccccll}
\tabletypesize{\scriptsize}
\tablecaption{Orbital parameters for the DSO  \label{tab2}}
\tablewidth{0pt}
\tablehead{
\colhead{$e$} & \colhead{$a$} & \colhead{$i$} & \colhead{$\Omega$}   &\colhead{ $\omega$} & \colhead{T}&\colhead{ P} \\ 
 \colhead{}  &\colhead{(mpc)} & \colhead{(deg.)} &\colhead{ (deg.)} &\colhead{ (deg.)} & \colhead{(yrs)} & \colhead{(yrs)} 
}
\startdata
  0.976$\pm$0.001 &
  33.0$\pm$3 &
  113$\pm$1 &
  76$\pm$8 &
  94$\pm$8 &
  2014.39$\pm$0.14 &
  262$\pm$38 \\
\enddata
\tablecomments{~The orbital parameters and their uncertainties have been derived on 
the basis of the UCLA measurements 
and the April 2014 data point we obtained using SINFONI at the VLT.
We assume a distance of 8~kpc and a black hole mass of 4$\times$10$^6$\solm.}
\label{elements}
\end{deluxetable}

\subsection{Interactions with the ambient medium}
\label{subsec:medinter}

If the DSO passes through an accretion wind from SgrA* it may develop a bowshock. 
In case it is indeed a dusty star, then one may expect to see cometary source structures 
quite similar to the sources X3 and X7 which are in the overall mini-cavity region 
just south of SgrA* at a projected distance of 0.8\arcsec and 3.4\arcsec \citep{muzic10}.
In mid-2014 the DSO is well within a sphere of hot gas surrounding Sgr~A* out 
to approximately the Bondi radius ($\approx 10^5 R_S$). 
As a dusty source, the DSO can therefore be regarded as an obvious probe for
strong winds possibly associated with SgrA*.
However,  there is no clearly resolved structure that can be considered as
a bowshock, although the DSO is already closer to SgrA* than X3 and X7.
This may indicate that the wind from SgrA* is
highly non-isotropic, possibly directed towards the mini-cavity
\citep{muzic10} and that the DSO has not yet passed through  that wind. 
However, the mass load of such a wind (due to the radiatively inefficient accretion mechanism) 
may not be high enough to allow for the formation of a prominent cometary tail structure.
The detailed density profile for the central region of the radiatively inefficient accretion flow 
is difficult to obtain. Methods are rather indirect and accretion model dependant
\citep{bag03, mar07, wang13}.
However, \citet{eck14IAU} have pointed out that the smaller size compared to 
X3 and X7 may be due to the higher particle density within the accretion stream close to
SgrA* \citep[e.g.][]{shc10}.

\subsection{Flare activity}
\label{subsec:flare}

A possibly efficient probe of the interaction of the DSO with its ambient environment
or with the black hole itself is monitoring the flux density originating from the central
few tenths of an arcsecond.
However, the results of these efforts have not been very revealing so far.

The NIR flare activity we observed through SINFONI during the peribothron approach in 2013/14
is in full agreement with the statistical expectations as we described them with \citet{witzel12}.
There was no exceptional activity, with 3 flares of a few milli-Jansky strength.

If the DSO would develop a bowshock while approaching the immediate environment of SgrA*
then this event might lead to shock accelerations of electrons and to correspondingly strong excursions in
the radio emission. However, the strength of these emission peaks depends critically on the size of the
bowshock and early estimates of the order of a 1-20~Jy in the decimeter to short centimeter wavelength 
range had to be revised to
values of the order of 0.01-0.2~Jy \citep{nar12, sad13a, cru13}. 
Despite a dense monitoring program with the VLA \citep{sjouwerman14}
strong radio flares have not yet 
been reported and the now predicted strength of the variability would be in the normal range of the
flux density variations observed towards SgrA* \citep[e.g.][]{markoff01, markoff07, eck12b}. 

So far in the X-ray observable $\ge2$keV band no elevated continuum flux density level or 
extraordinary X-ray variability has been reported \citep{haggard2014}.
Such an extra emission would have been expected to originate from the shock-heated gas \citep{gil12}.

Although SgrA* is extremely faint in the X-ray bands, 
it is strongly variable in this domain of the electromagnetic spectrum
\citep{bag03, bag01, por03, por08, eck12b, nowak12, degenaar2013, barr14, mossoux14,neilsen13}.
The statistical investigation of the near-infrared variability by \citet{witzel12} 
suggests that the past strong X-ray variations
are potentially linked with the origin of the observed X-ray echos
\citep{rev04, sun98, ter10, capelli12}.
Assuming an underlying Synchrotron Self Compton (SSC) process
the NIR variability can in fact explain the required X-ray flare fluxes as natural and non-exceptional 
phenomenon of the source.
Therefore, SgrA* is the ideal extremely low
accretion rate target that allows us to study this particular phase 
in which apparently most super massive black holes spend their
lifetime.
Phenomena like the passage of the DSO may dominate the variability of objects in this phase 
throughout the electromagnetic spectrum.

While the DSO is a very compact continuum and line emitting source 
(see section \ref{subsec:redpost} and \citet{eck13}) its peribothron distance is rather small 
(see section \ref{subsec:orbit}).
Hence it is still an open question if and when some activity of SgrA* is triggered by the DSO fly-by.

\subsection{The DSO as a young accreting star}
\label{subsec:ttau}

Large line widths are common amongst
pre-main sequence stars including both T~Tauri stars (with an age of about $10^5-10^6$ years) and proto-stars 
(with an age of about $10^4-10^5$ years) with an infalling envelope that forms a disk close 
to the star.
Bertout (1994) already pointed out that
Doppler broadening from pre-main sequence stars may range roughly from 50-500 km/s in the course of the accretion phase.
As an example, hydrogen recombination and Na~D line profiles of several 100 km/s in a number of pre-main sequence stars
(e.g. T~Tau, DG~Tau, DR~Tau, AS205, SCrA) are shown.
The M0V classical T Tauris star LkCa-8 (IP Tau) \citep{wolk96, motooka13}
has a 600-700 km/s Br$\gamma$ linewidth \citep{edwards13} 
quite comparable to the width currently found for the DSO. Another case of a low mass star with
exceptionally large line widths is DK~Tau~A with an 800 km/s wide line \citep{eisner07}. 
It is listed by \citet{herczeg14} as a K8.5 star with a mass of 0.68~\solm.

Without doubt the Br$\gamma$ line traces high excitation regions, however, in the case of
young embedded proto stars it is currently unclear, whether these regions are associated
with accretion funnel flows, the jet base \citep{davis11}
or less collimated ionized winds. All of these elements can contribute to the emission and the large
observed line width. In the case of the DSO there are several mechanisms that can contribute
to a large line width.

$i)$ {\it Contribution from collisional ionization in a bowshock:}
A possible origin of a broad wide Br$\gamma$ emission line was discussed by \citet{sco13} 
on the basis of the bowshock
model that is relevant for the supersonic motion of the object through the hot ambient
ISM emitting X-rays. They show that Br$\gamma$ emission may arise from the collisional
ionization and the gas cooling in the narrow but dense cold 
($\sim 10^5$--$10^6\,\rm{K}$) and shocked layer of the stellar wind. 
The high densities ($\sim 10^8\,\rm{cm^{-3}}$) in this layer can explain the observed emission measure.

$ii)$ {\it Contribution from wind drag in a bowshock:}
The large increase in FWHM line width from 137 km/s in 2006 to 730 km/s in 2014
could also be related to the increase in orbital velocity from about 1200 km/s to
almost 9000 km/s at peribothron . Discussing the emission from photoionized stellar wind
bow shocks \citet{canto05} calculate the change of velocity in the thin shocked layer
that develops while the source is moving through the ISM. In the context of the DSO
this effect has not yet been discussed before. In their equations (19) and (33) they
approximate the dependence of that velocity as 
$v_{sl} \propto v_w \times f(v_a, R, \theta, \phi)$. 
Here $v_{sl}$ is the velocity in the shock layer, 
$v_w$ is the stellar wind velocity, 
$v_a$ is the velocity relative to the ISM. 
The radius R and the angles $\theta$ and $\phi$ describe the geometry of the shock
front. It is the change of $v_{sl}$ across the shock
front that may contribute to the observed Br$\gamma$ line width.
The analytic solution of \citet{wilkin96} for the thin steady-state bow-shock 
layer yields the estimate for the shock-layer velocity 
$v_{\rm{sl}}\approx 2 v_{\rm{a}}\theta/[3(1+v_{\rm{a}}/v_{\rm{w}})]$ 
close to the symmetry axis, 
where the angle $\theta$ is small. 
The ratio of this velocity at the same $\theta$, but different epochs, 2006 and 
the peribothron crossing, yields $v_{\rm{sl}}^{\rm{per}}/v_{\rm{sl}}^{2006} \approx 1.07$--$1.26$ 
for the terminal wind velocities of $100$--$400\,\rm{km/s}$, respectively. 
Thus, the increase in velocity by about $10\%$ could by contributed by wind drag in a bow-shock layer.

$iii)$ {\it Contribution from stellar or disk winds:}
There can also be a contribution to Br$\gamma$ emission from the gaseous inner disk, stellar
wind, stellar-field driven wind (X-wind) or disk wind \citep{lima10}
that can originate from the corotation radius to several astronomical units 
(see \citet{kraus08}, for discussion and their Fig.~1). 
\citet{guenther11} shows that for classical pre-main sequence
stars, wind velocities of a few hundred km/s can occur.

$iv)$ {\it Tidal contribution:}
The increase in FWHM of DSO would then be caused by the tidal stretching and perturbation of 
the accretion disk, especially close to the peribothron, which would consequently lead to 
larger velocity dispersions of inflow and outflow streams. 
Simple considerations analogous to the computation of tidal compression presented by 
\citet{jalali14} show the increase of velocity deviation. 
There are several ways to assess the importance
of tidal stretching of the DSO along its orbit from our data:
\\
{\it $\alpha$)~}
If the total pre-peribothron linewidth of about 720~km/s was dominated by tidal stretching 
then a minimum source size of about 65~mas is expected based on the mean slope of 1000 km/s 
over a projected orbital path of about 90~mas (i.e. $\sim$11~km/s/mas) within the past two years.
\\
{\it $\beta$)~}
Attributing the 2008-2013 size estimate of 42$\pm$10~mas \citet{gil13b}
to the year 2013 and assuming a free gas cloud subjected to orbital stretching along the 
orbit we find that the source should be 5-8 times larger, i.e 210~mas to 336~mas, close to peribothron.
We cannot confirm such a large size from our Br$\gamma$ line maps in
Fig.~\ref{fig:linecentroids} (see also Tab.~\ref{tab:linecentroids}).
\\
{\it $\gamma$)~} The separation of apparently simultaneously observed extreme velocity components of G2
close to the peribothron passage (Fig.1 and Fig.15 in \citet{gil13b, pfu14b} which is consistent with a cut 
through their pv-diagram in Fig.1) implies  a  size between 90~mas and 150~mas along the orbit.
With our data we only see a single lined DSO either red- or blue-shifted with a diameter of $<$20~mas 
and, given the low foreshortening (section~\ref{subsec:orbit} and Fig.~\ref{fig:fore}),
we cannot confirm the presence of multiple sources or a large source extent 
(see section~\ref{subsec:redpost} and Fig.~\ref{fig:linecentroids}).
\\
{\it $\delta$)~}
We measured very close to the points at which extreme orbital velocities in the
red and blue can be observed. At these positions for an extended tidally stretched source 
the emission previously blue of the source center will become red-shifted and blue-shifted, respectively.
Hence a linewidth  that can be up to a factor 2 narrower is expected. However, the post-peribothron linewidth
is about a thrird of the pre-peribothron value, and the small source sized are in conflict with
an extended tidally stretched source.
In addition the orbital compression extected for such a scenario would imply a larger line width and a 
higher line flux density due to the increased dnesity of the emitting gas volume.
Instead for an dust-enshrouded accreting stellar object line variability in integrated line flux density 
and line shape is expected.

$v)$ {\it Contributions from accretion:}
However, there can also be a contribution from the gas accretion of the circumstellar envelope
onto the stellar surface if the DSO is a young stellar object as has already been
proposed and discussed \citep{mur12,eck13,sco13,zajacek14,2014ApJ...789L..33D}. 
In this framework an increase of the Br$\gamma$ line width as a function of time 
could result from an increased perturbation of the envelope or disk that 
leads to an enhanced velocity dispersion in the accretion stream onto the central star
as it gets closer to the peribothron. 

In the following we investigate
if the infalling gas that is approximately in free-fall and is being shocked upon reaching
the stellar surface can explain the observed large linewidth of the DSO Br$\gamma$ line, 
that evolved from FWHM$(\rm{Br}\gamma) \sim 200\,\rm{\kms}$ in 2006
to FWHM$(\rm{Br}\gamma) \sim 700\,\rm{\kms}$ in 2014
as laid out in section~\ref{subsubsec:brg} \citep[see also,][]{phi13,gil13b}. 
This corresponds to radial velocities $v_{\rm{r}}$ of infalling material that range from about 100 $\rm{\kms}$ 
to several hundred $\rm{\kms}$.

\subsubsection{The model geometry}
\label{subsubsec:ttaumodel}

For simplicity, we  consider an \textit{axisymmetric magnetospheric accretion model} \citep[see][for review]{2007prpl.conf..479B} for the accretion on pre-main sequence stars where the gas moves ballistically along the magnetic field lines from the innermost orbit of the disk and gains large infall velocities of the order of $\sim 100\,\rm{km s^{-1}}$ \citep{1994ApJ...426..669H}. 
Unlike the boundary layer model, the magnetospheric accretion scenario can indeed explain
observed redshifted absorption minima at free-fall velocity and blue-ward asymmetry in emission lines
\citep[][and references therein]{1998ApJ...492..743M}.

\begin{figure}[hbt!]
\centering
\includegraphics[width=0.99\linewidth]{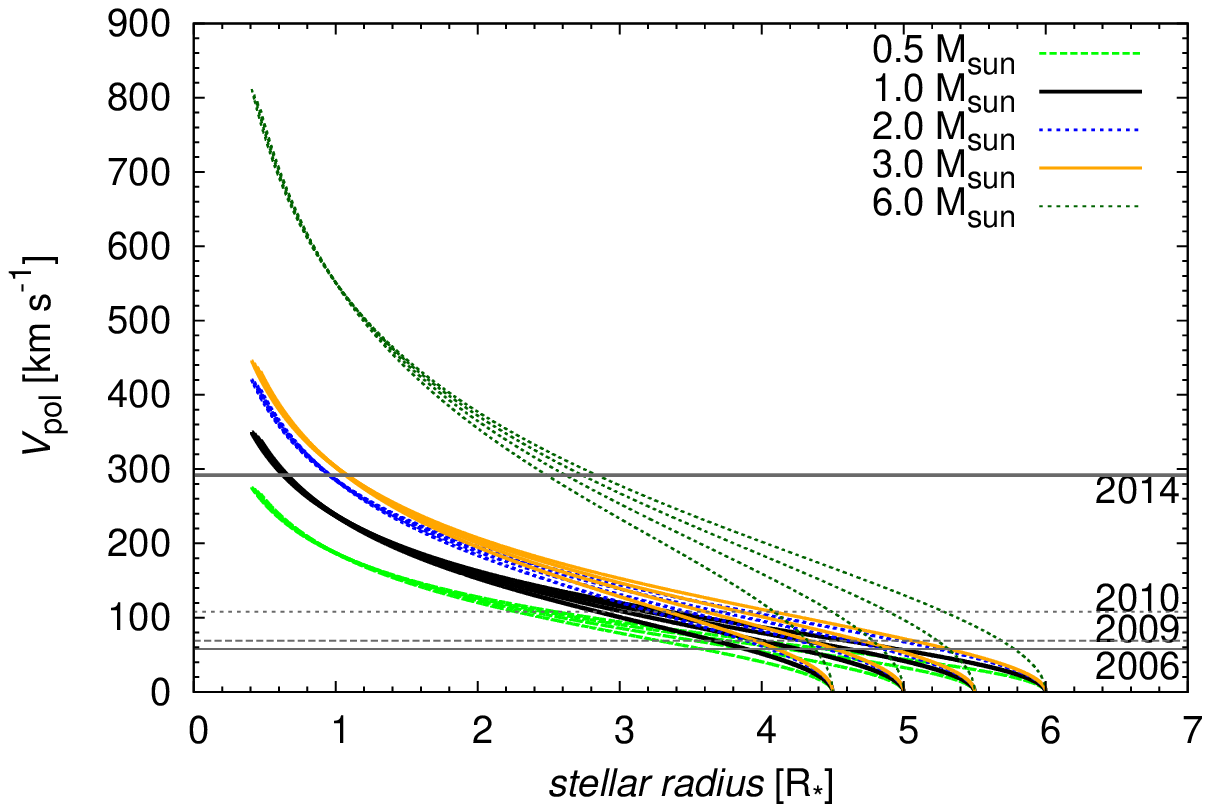}
   \caption{The poloidal velocity profile as a function of the distance from the pre-main sequence star in a magnetospherical accretion model \citep{1994ApJ...426..669H} for different masses of pre-main sequence stars at $\sim 1 \,\rm{Myr}$ \citep{Siess2000}. The gray horizontal lines represent the observed radial velocity for years 2006, 2009, 2010, and 2014 with an increasing tendency \citep[][this work]{phi13,gil13b}.}
   \label{fig_velprof}
 \end{figure}

\begin{figure}[hbt!]
\centering
\includegraphics[width=0.99\linewidth]{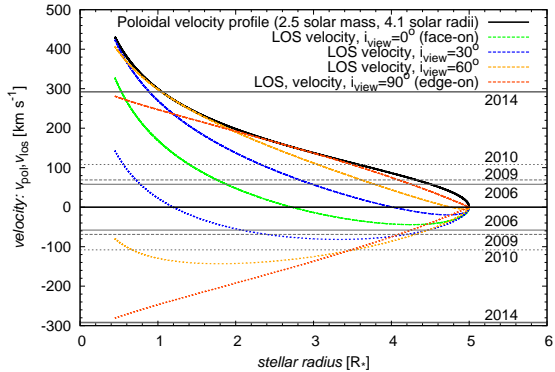}
   \caption{The profiles of maximum line-of-sight velocity as a function of the distance from the star in a magnetospherical 
accretion model \citep{1994ApJ...426..669H} for $2.5\,M_{\odot}$ pre-main sequence star for a different inclination of the 
view of emerging radiation. The receding flow (nearer to the observer, mostly red-shifted, positive velocities) is labeled by dashes, 
the approaching flow (further away, mostly blue-shifted, negative velocities) is represented by dots. 
The gray horizontal lines represent the observed radial velocity  $v\sin{i}$ 
for years 2006, 2009, 2010, and 2014 with an increasing tendency \citep[][this work]{phi13,gil13b}.}
   \label{fig_velprof_los}
 \end{figure}        
 
 The presence of a magnetic field around pre-main sequence stars is justified by the observation of the Zeeman broadening of photospheric lines \citep{1999ApJ...516..900J,2001ASPC..248..527J} as well as by the measurement of the electron cyclotron maser emission \citep{2003A&A...406..957S}. The inferred field strength is $\sim 1$--$3\,\rm{kG}$. In the context of the dipole magneto-accretion model, in which the gas is in free-fall, the truncation radius in terms of stellar radii is \citep[e.g.,][]{2007prpl.conf..479B}:
 
\begin{equation}
\frac{R_{\rm{T}}}{R_{\star}} \approx 6.5 B_{3}^{4/7} R_{2}^{5/7} \dot{M}_{-8}^{-2/7} M_{1}^{-1/7}\,,
\label{eq_truncation_radius}
\end{equation}   

where the strength of the dipole magnetic field at the equator $B_{3}$ is in $\rm{kG}$, the stellar radius $R_{2}$ is in units of $2R_{\odot}$, the accretion rate $\dot{M}_{-8}$ is expressed in $10^{-8}M_{\odot}\rm{yr}^{-1}$, and the stellar mass $M_{1}$ is in units of $1 M_{\odot}$. The truncation radius in Eq. \ref{eq_truncation_radius} is derived for gas in free-fall in the spherical symmetry. For disk accretion it may serve as an upper limit, since the ram gas pressure is higher in that case and the truncation radius is thus shifted inwards. 

For stable accretion to proceed, the truncation radius expressed by Eq. \ref{eq_truncation_radius} has to be smaller than the corotation radius $R_{\rm{co}}$, $R_{\rm{T}}\lesssim R_{\rm{co}}$, at which the Keplerian angular velocity is equal to the rotational angular velocity of the star,
\begin{equation}
R_{\rm{co}}\approx 4.2\, M_1^{1/3}P_{1}^{2/3}\,R_{\odot}, 
\end{equation}
where $M_{1}$ is the stellar mass in units of $1\,M_{\odot}$ and $P_{1}$ is the stellar rotation period in units of $1$ day \citep[see][for discussion]{2007prpl.conf..479B}.
   The inner portion of the disk is purely made up of gas up to the dust sublimation radius, which according to simulations by \citet{2004ApJ...617.1177W}, may be expressed in terms of the dust sublimation temperature and the stellar effective temperature as,

\begin{equation}
R_{\rm{sub}}=R_{\star}\left(\frac{T_{\rm{sub}}}{T_{\star}}\right)^{-2.085}\,,
\end{equation}  
which for typical values of $T_{\star}$ (spectral types K, M: $3000$-$4500\,\rm{K}$) and $T_{\rm{sub}}\approx 1500\,\rm{K}$ has values of $\sim 4$--$10\,R_{\star}$. Beyond the dust sublimation radius dust can coexist with the gaseous phase. The emerging radiation from the accretion flow is reprocessed by the circumstellar dust, giving  rise to the infrared excess. 

For the calculation of the velocity profiles of the accretion flow we assume the truncation radius to be at $R_{\rm{T}}=5\,R_{\star}$ \citep{1998ApJ...492..323G,2014A&A...561A...2A}, which is close to the estimate in Eq. \ref{eq_truncation_radius}. 
We compute the poloidal velocity profiles in the framework of the magnetospheric accretion model 
\citep[see Eq.1 and 3 of][]{1994ApJ...426..669H} 
for $0.5$, $1.0$, $2.0$, and $3.0\,$ $M_{\odot}$ pre-main sequence stars with a stellar radius of 
$2.1$, $2.6$, $3.6$, and $4.8$ $R_{\odot}$, respectively, at 1~Myr \citep{Siess2000} for solar metalicity and no overshooting. 
See Fig. \ref{fig_velprof} for the comparison of the poloidal velocity for the observed radial velocity, which is observed to increase with the approach of the DSO to the peribothron.  For earlier epochs the observed FWHM is consistent with the accretion onto a low-mass object of $\sim 0.5$--$1\,M_{\odot}$. 
To explain the higher FWHM in 2014 a massive pre-main sequence star of Herbig Ae/Be type is needed at the first glance, 
since for lower-mass stars only the upstream parts reach comparable velocities. See the poloidal velocity profile of 
$6 M_{\odot}$ star with the radius of $2.9 R_{\odot}$ in Fig.~\ref{fig_velprof}.
However, such a massive stellar core having a luminosity of $\gtrsim 100 L_{\odot}$ is inconsistent with the luminosity 
constraint on the DSO ($\le$10 \solar) and the pre-main sequence stage is also very short \citep{Siess2000}.

The physics of circumstellar material of pre-main sequence stars is generally more complex, especially close to the 
SMBH, where the disk surrounding the star is expected to be warped and perturbed by tidal effects. 
Although basic observational signatures of pre-main sequence stars (strong stellar magnetic fields, truncation radius, accretion shocks observed mainly for classical pre-main sequence stars) are in accordance with the magnetospheric accretion model \citep{2007prpl.conf..479B} and suggest that Br$\gamma$ originates in gas infall rather than outflow \citep{1996ApJ...456..292N}, it is plausible that there is a contribution to Br$\gamma$ emission from stellar or disk winds

We note that the line profile may be generally non-symmetric and its width dependent on the inclination, at which the emerging emission is viewed. This is demonstrated by the profiles of the maximum line-of-sight velocity in Fig. \ref{fig_velprof_los}, where we plot separately approaching  (mostly blue-shifted) and receding (mostly red-shifted) accretion streams for $2.5\,M_{\odot}$ pre-main sequence stars
\citep[at $\sim 1\,\rm{Myr}$,][]{Siess2000}. 
The separation between the dashed and the dotted lines for each stellar radius and inclination is an approximate measure of the observed
linewidth.
The line broadening is generally bigger for a larger inclination \citep[see also][for detailed radiative transfer modeling]{1998ApJ...492..743M}. 
The results in Fig. \ref{fig_velprof_los} clearly show that with a star of about $2\,M_{\odot}$  the observed Br$\gamma$ line widths,
covering the full range from about 200 $\kms$ to 700 $\kms$ can be reproduced.
From the range of maximum and minimum velocities it is also evident that the line profile can be asymmetric and skewed to one side.

\subsubsection{Accretion luminosity and rate}
\label{subsubsec:ttauLM}

  In fact, the broad hydrogen Br$\gamma$ line with the full width at half maximum of the order of $\sim~100\,\rm{km\,s^{-1}}$ is frequently observed in the spectra of accreting pre-main sequence stars \citep[detection rate $70\%$--$74\%$,][]{2001A&A...365...90F,2014arXiv1409.4897I} and appears to be a useful tracer of magnetospheric accretion on embedded pre-main sequence stars \citep{1998AJ....116.2965M,1998ApJ...492..743M,2001A&A...365...90F,2004AJ....128.1294C}. 
Here the star is assumed to accrete matter form its envelope or the inner edge of an accretion disk.
Accretion from the surrounding interstellar medium can be considered as insignificant.
 
 The correlation between Br$\gamma$ emission-line luminosity and accretion luminosity is found to be tight \citep{1998AJ....116.2965M,2004AJ....128.1294C}. The empirical relation between emission line and accretion luminosities is based on various signatures of accretion luminosity (H$\alpha$ luminosity, optical and UV excess). The recent fit is as follows \citep{2014A&A...561A...2A},

 \begin{equation}
 \log{(L_{\rm{acc}}/L_{\odot})}=\zeta_1\log{[L(\rm{Br} \gamma) / L_{\odot}]} + \zeta_2~.
 \label{eq_correlation_luminosities}
 \end{equation}
 
with $\zeta_1$=1.16$\pm$0.07 and $\zeta_2$= 3.60$\pm$0.38.
 This correlation may then be extended to heavily extincted proto-stars that are enshrouded in a dusty envelope.
  
   If we naively apply this relation to the DSO and its Br$\gamma$ emission-line luminosity of $L(Br\gamma)= f_{\rm{acc}} \times 10^{-3}\, L_{\odot}$, where $f_{\rm{acc}}$ is a factor of the order of unity, we get a reasonable range for the accretion luminosity, $\log{(L_{\rm{acc}}/L_{\odot})} \approx 1.16 \log{f_{\rm{acc}}}+0.12$; $L_{\rm{acc}}=1.3 \times 14.5^{\log{f_{\rm{acc}}}}\,L_{\odot}$, 
and for $f_{\rm{acc}}=\{1,2,3,4\}$
yielding $(1.3,3.0,4.7,6.6)\,L_{\odot}$. 
   
     For the assumption of the innermost radius of $R_{\rm{in}}=5\,R_{\star}$, the accretion rate is given by \citep{1998ApJ...492..323G}
   
   \begin{equation}
  \dot{M}_{\rm{acc}} \cong \frac{L_{\rm{acc}}R_{\star}}{GM_{\star}} \left(1-\frac{R_{\star}}{R_{\rm{T}}}\right)^{-1}\\
   \label{eq_mass_accretion_rate-0}
   \end{equation}

\noindent
which can be written as

   \begin{equation}
  \dot{M}_{\rm{acc}}  \approx \xi \left(\frac{L_{\rm{acc}}}{L_{\odot}}\right)\left(\frac{R_{\rm{\star}}}{R_{\odot}}\right)\left(\frac{M_{\rm{\star}}}{M_{\odot}}\right)^{-1}\,M_{\odot}\rm{yr^{-1}}
   \label{eq_mass_accretion_rate}
   \end{equation}
with $\xi=4.1 \times 10^{-8}$.

  Inserting the estimated values for mass, radius, and the accretion luminosity, we obtain an accretion rate of the order of $\lesssim 10^{-7}\,M_{\odot}\,\rm{yr^{-1}}$, which is about ten times larger than the median value observed for pre-main sequence stars in Taurus and Chameleon~I regions \citep{1998ApJ...495..385H}. It is, however, consistent with the span of pre-main sequence accretion rates, which seem to evolve with the age of the pre-main sequence star as $\dot{M}_{\rm{acc}} \propto t^{-2.1}$ \citep{2008ApJ...689..308B}. 
   
   The gas outflow rate was shown to correlate with the accretion in pre-main sequence systems. The ratio of rates was established approximately as $\dot{M}_{\rm{w}}/\dot{M}_{\rm{acc}}\sim 0.1$ \citep[][and references therein]{2006ApJ...646..319E}, which corresponds to the order of $\dot{M}_{\rm{w}}\lesssim 10^{-8}\,M_{\odot}\,\rm{yr^{-1}}$. This order of magnitude for the wind outflow rate was discussed by \citet{sco13} for the wind-wind bow-shock origin of Br$\gamma$ emission.  
   
   The estimates of accretion luminosity and accretion and mass-loss rates are upper limits since there may be contribution to Br$\gamma$ flux from other sources than accretion flows, namely stellar wind or disk outflows \citep{kraus08}. 
  
\subsubsection{Density and emission measure}
\label{subsubsec:ttaudens}

\begin{figure}[hbt!]
\centering
\includegraphics[width=0.99\linewidth]{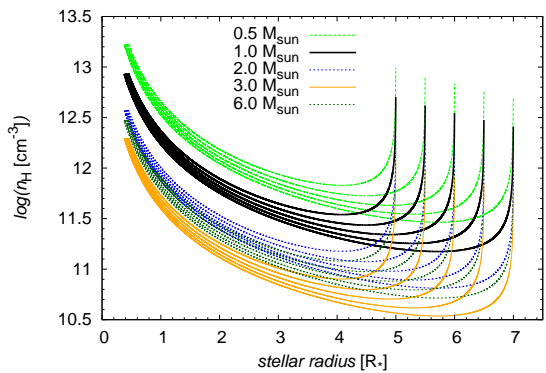}
\caption{The radial number density profile as a function of the distance from the star in units of stellar radii. The computed profile is valid for the steady axisymmetric accretion on a pre-main sequence star. Individual lines correspond to accretion flows along magnetic streamlines for a particular mass of a star (see the labels). The radius of a star is adopted from the evolutionary tracks of \citet{Siess2000} at $\sim 1\,\rm{Myr}$.}
\label{img_denprof}
\end{figure}

  The radial density profile of the accretion flow may be inferred based on the estimated values of pre-main sequence star mass, radius, mass-accretion rate and the assumed size of the magnetosphere. Assuming an axisymmetric steady flow of matter along the streamlines, the following relation holds for the hydrogen number density \citep{1994ApJ...426..669H},

  \begin{equation}
  n_{\rm{H}}(r)=\frac{\dot{M}_{\rm{acc}}}{4\pi m_{\rm{H}}(\frac{1}{r_{\rm{mi}}}-\frac{1}{r_{\rm{mo}}})}\frac{r^{-5/2}}{(2GM_{\star})^{1/2}}\frac{(4-3y)^{1/2}}{(1-y)^{1/2}}\,,
  \label{eq_density_profile}
  \end{equation}

where the magnetic streamlines are described by $r=r_{\rm{m}}\sin^2{\theta}$, where $\theta$ denotes the angle between the magnetic dipole axis and the radius-vector $\mathbf{r}$; in Eq. \ref{eq_density_profile} $y=r/r_{\rm{m}}=\sin^2{\theta}$ and $r_{\rm{mi}}$ and $r_{\rm{mo}}$ stand for the radius of the innermost and the outermost streamline intersecting the accretion disk, respectively; we take $r_{\rm{mi}}=5\,R_{\star}$ and $r_{\rm{mo}}=7\,R_{\star}$ for definiteness. The mass accretion rate is held fixed at $\dot{M}_{\rm{acc}}=10^{-7}\,M_{\odot}\rm{yr^{-1}}$ in accordance with Eq. \ref{eq_mass_accretion_rate}. The density profiles for the same set of stars as in Fig. \ref{fig_velprof} are plotted in Fig. \ref{img_denprof}.

\begin{figure}[hbt!]
\centering
\includegraphics[width=0.99\linewidth]{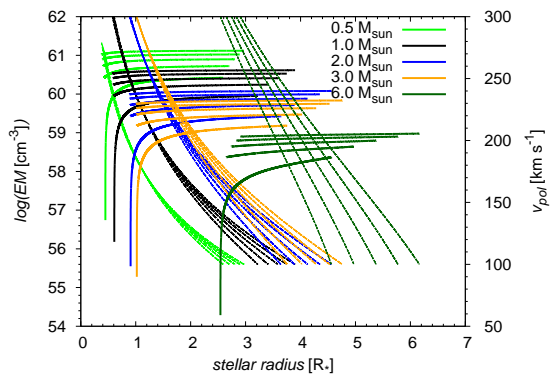}
\caption{The logarithm of emission measure EM 
(solid lines, with labels on the left vertical axis) and poloidal velocity profile $v_{\rm{pol}}$ (dot-dashed lines, with labels on the right vertical axis) as a function of the distance from the star in units of stellar radii. 
The computed profile is valid for the steady axisymmetric accretion on a pre-main sequence star. Individual lines correspond to accretion flows along magnetic streamlines for a particular mass of a star (see the key). The radius of a star is adopted from the evolutionary tracks of \citet{Siess2000} at $\sim 1\,\rm{Myr}$.}
\label{img_emprof}
\end{figure}

The density profile in Fig. \ref{img_denprof} enables to estimate the emission measure, EM~$\propto n^2_{\rm{e}} V$, under the assumption $n_{\rm{e}} \approx n_{\rm{H}}$, see Eq. \ref{eq_density_profile}. 
The computation is performed for the distance range where the poloidal velocity, Fig. \ref{fig_velprof}, reaches the values of $v_{\rm{pol}}=200\pm 100\,\rm{km s^{-1}}$, which roughly corresponds to the observed FWHM of the Br$\gamma$ line.  Finally, we get the profiles of cumulative emission measure for a different mass of a pre-main sequence star according to the evolutionary tracks by \citet{Siess2000} at $\sim 1\,\rm{Myr}$, see Fig. \ref{img_emprof}. 
The emission measure is of the order of $10^{58}$--$10^{61}\,\rm{cm^{-3}}$, being higher for lower-mass stars, 
and these values originate from close to the star, on the scale of $\sim 1$--$3\,R_{\star}$. 
This implies that luminous line emission at the observed high velocities can originate from close to a few solar mass star.

One should consider these values highly estimative because of the uncertain values of mass-accretion rate and the size and character of the magnetosphere. The temperature of the accretion flow was also not discussed. However, models of the accretion on pre-main sequence stars show that the infalling gas is shock-heated and the Br$\gamma$ line can be effectively produced close to the stellar surface \citep{2007prpl.conf..479B}. In principle it is possible to reproduce the emission measure of $\sim 10^{57}\,\rm{cm^{-3}}$ that is obtained in the cold bow-shock model by \citet{sco13} and that corresponds to the observed flux of Br$\gamma$ emission. Thus both mechanisms, wind-wind interaction and gas infall, can contribute in case the DSO is a young stellar object.

Given the accretion rate of $\lesssim 10^{-7} M_{\odot}\rm{yr^{-1}}$ the star associated with the DSO would be 
embedded within the hot accretion flow surrounding the star with a probably complex geometry. 
On the length-scale of one stellar radius the density profiles 
in Fig. 10 imply large infrared K-band and visible extinction of 
$A(K) \sim 0.1 \times A(V) \approx 0.1 \times (1.8 \times 10^{21})^{-1}\int n_{\rm{e}}(l) \mathrm{d}l$, 
which is $\sim 59$, $37$, $20$, and $16$ magnitudes for a
$0.5\,M_{\odot}$, $1.0\,M_{\odot}$, $2.0\,M_{\odot}$, and $3.0\,M_{\odot}$ star, respectively.
Combined with the possible contribution of an extended outer dust shell and a warped or inflated outer disk,
this is plenty of extinction to dim the light from the central star and produce the observed continuum 
characteristics of the DSO  \citep{eck14, eck14IAU, eck13, phi13, gil12},

The material within the accretion flow is certainly not homogeneous. To first order we assume that
it consists of cloudlets, sheets or filaments that have a dense, optically thick core surrounded 
by a shell of optically thin material.
Since the overall line emission is dominated by optically thin material and
given the uncertainties in the Pa$\alpha$/Br$\gamma$ and He~I/Br$\gamma$ line ratios
the maximum contribution of the optically thick material to the line fluxes can only be of the order of 0.05.
Since the dense material is bright the
ratio $\eta$ between the mean emission measure of the optically thin and thick material needs to be involved.
Hence a volume filling factor around 0.05$\times$$\eta$ will reproduce the optically thin
line ratios and give a substantial contribution to the observed velocity profile.
In addition, there will be a temperature gradient from the start of the accretion stream near the
dust sublimation radius and the contact point on the stellar surface, leading to an enhanced contribution of the
higher velocity material.

      The observed accreting pre-main sequence stars generally have a lower accretion rate than the time-averaged infall rate \citep{2014arXiv1401.3368A}. As a result, the gas from an infalling envelope is thought to be accumulated first in the quasi-Keplerian circumstellar disk. The accumulation of matter continues until the instability causes an increase in the mass transfer by about three--four orders of magnitude from the disk to the star, the so-called episodic accretion \citep{2014arXiv1401.3368A}.  If the DSO is indeed an embedded accreting pre-main sequence star, the tidal effects from the SMBH lead to a gravitational instability which, combined with magneto-rotational instability \citep{2009ApJ...694.1045Z}, can cause a continual mass transfer from the disk, especially close to the peribothron where the tidal radius shrinks to $\lesssim 1\,\rm{AU}$ for a $1 M_{\odot}$ star, see also the tidal radius discussion related to Fig. \ref{fig_tidal_radius}. 
   
   Let us sum up that the observed emission up to now is not in contradiction with the scenario of a pre-main sequence star that is surrounded by a dusty envelope and accretes matter from an accretion disk inside the dust sublimation radius. 
Hot accretion flows as discussed here, possibly combined with disk winds \citep{guenther11},
can indeed produce emission lines with FWHM of several hundred km/s.  
Hence we find that for a 1-2~\solm ~embedded pre-main sequence star these two effects can already fully account for the observed Br$\gamma$ line widths.
However, in general the observed Br$\gamma$ line profile and flux may result from the combination of hydrogen recombination emission of the gaseous-dusty envelope photoionized by nearby stars \citep{shch14}, collisionally ionized cold bow-shock layer \citep{sco13}, and the hot accretion flow on a pre-main sequence star, as is discussed here. If and to what extent each of these processes contributes to the final emission will be constrained by further observations and modeling during the post-peribothron phase.

\section{Possible origin, stability and fate of the DSO}
\label{sec:origin}

There is evidence of both young and more evolved stars in the Galactic centre 
\citep{gen10} that lie in the sphere of influence ($\sim 2\,\rm{pc}$) of the 
super massive black hole (SMBH). Mutual interactions among stars cause the oscillations 
of their orbital eccentricity via the mechanism of resonant relaxation 
\citep*{hop06} or the Kozai oscillations \citep{kar07, chen2014}. 
These can set some stars on a plunging trajectory towards the SMBH \citep{zajacek14}.
Similarly, with \citet{jalali14} we have shown that young stellar objects can efficiently be formed
on plunging orbits in the vicinity of super massive black holes as a consequence of orbital compressing 
of infalling gas clumps. 
An embedded young star/protostar is surrounded by an accretion disk whose orbit orientation 
can be any (direct, retrograde or perpendicular) with respect to the orbit of the host star 
around the SMBH. These dusty S-cluster objects \citep[DSOs;][]{eck13} have an infrared excess and the 
currently observed DSO may indeed serve as a paradigm of these objects.

The restricted three-body problem may be used to obtain the approximations for critical stability (Hill) radii of disks. Using the restricted circular three-body problem the equation of motion for a mass element in the rotating frame of star--SMBH becomes \citep[i.e.,][]{inn79,inn80}:

\begin{equation}
\frac{\mathrm{d}^2 r}{\mathrm{d}t^2} \simeq \left(\Omega^2-\frac{\mathrm{d}^2 V}{\mathrm{d}R^2}-\frac{GM_{\star}}{r^3}\right)r \pm 2\Omega v_{\rm{r}}\,,
\label{eq_motion} 
\end{equation} 
where $r$ is the distance of a mass element from the star and $R$ labels the distance of the star from the SMBH, $r\ll R$. For $R\gg R_{\rm{g}}$, the gravitational potential of the black hole is approximately equal to the Newtonian $V\equiv V(R)$, hence $-\mathrm{d}^2V/\mathrm{d}R^2=2GM_{\bullet}/R^3$. The angular frequency of the circular motion for the star is $\Omega^2=GM_{\bullet}/R^3$ and for the minor body $\omega^2=GM_{\star}/r^3$. The difference between direct and retrograde orbits arises from the different signs of the Coriolis term $\pm 2\Omega v_{\rm{r}}$, where $v_{\rm{r}}=\omega r$. 

\begin{figure}[!htb]
\centering
\includegraphics[width=0.99\linewidth]{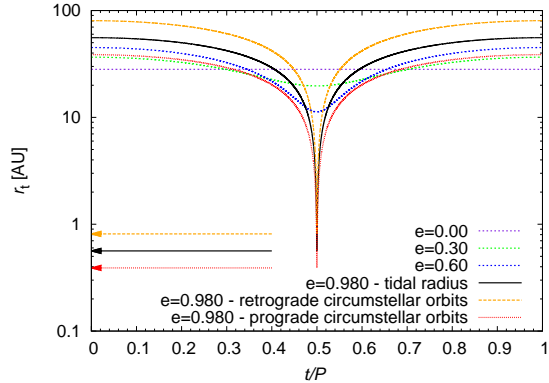}
\caption{
Temporal evolution of the tidal Hill radius for the orbits of a 1\solm ~star around the SMBH 
with a different orbital eccentricity.
At $t/P=0.5$ the peri-center passage occurs. 
For the inferred orbit of the DSO, tidal radii for prograde as well as retrograde orbits are depicted. 
Actual values of the tidal radii at the peribothron are depicted by corresponding arrows. 
The current orbital solution implies a considerable tidal stripping for distances from the star $\gtrsim 1\,\rm{AU}$.}
\label{fig_tidal_radius}
\end{figure}

When generalized for orbits with eccentricity $e$, one gets the following ratio of critical tidal radii for retrograde and direct disks, $r_{\rm{H,r}}$ and $r_{\rm{H,d}}$, respectively \citep{inn79}:
\begin{equation}
\frac{r_{\rm{H,r}}}{r_{\rm{H,d}}}=\left[\frac{5+e+2(4+e)^{1/2}}{3+e}\right]^{2/3}\,,
\label{eq_ratio}
\end{equation}
In terms of the critical tidal radius, 
\begin{equation}
r_{\rm{t}}=R(t)(M_{\star}/(3M_{\bullet}))^{1/3}~, 
\label{eq:hill}
\end{equation}
the critical Hill radius for prograde orbits $r_{\rm{d}}$ and retrograde orbits $r_{\rm{r}}$ may be expressed as 
$r_{\rm{r}}=3^{1/3}r_{\rm{t}}$ and $r_{\rm{d}}=3^{-1/3}r_{\rm{t}}$. 
Since the mass of the SMBH  $M_{\bullet}$ is larger by at least 5 orders of magnitude 
than the mass of any star $M_{\star}$, the tidal Hill radius $r_{\rm{t}}$ expressed by eq.~\ref{eq:hill}
extends up to Lagrangian points L1 and L2, beyond which the circumstellar matter is strongly 
tidally perturbed and may escape the Roche lobe of the star 
\citep[see Fig.18 by][]{eck13}.
The temporal evolution of the tidal radii for the current orbital elements and the stellar mass of 
$m_{\star}=1\,M_{\odot}$ is shown in Fig. \ref{fig_tidal_radius}. 

\begin{figure}[!htb]
\centering
\includegraphics[width=0.99\linewidth]{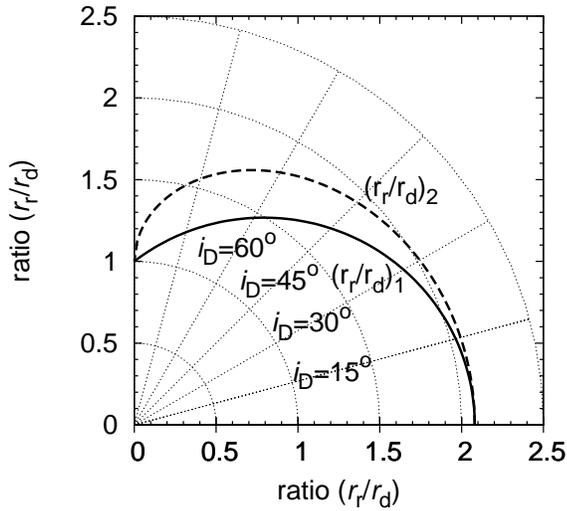}
\caption{
The polar plot of the dependence of the ratio of critical tidal radii of retrograde ($r_{\rm{r}}$) and
prograde ($r_{\rm{d}}$) disks on the inclination of the accretion disk around a pre-main 
sequence star with respect to the orbital plane of the star around the SMBH. 
In this plot the vertical and horizontal axis represent the same quantity - 
the ratio between retrograde and prograde tidal radii. 
The ratio dependence on the inclination was derived by \citet{inn80}. 
The plotted relations have the form of 
$(r_{\rm{r}}/r_{\rm{d}})_{1,2}=\{1+f_{1,2}/2+[f_{1,2}+(f_{1,2}/2)^2]^{1/2}\}^{2/3}$, 
where $f_1=4(\cos^2{i})/3$ and $f_2=4\cos{i}/(2+\cos{i})$. 
The relation $(r_{\rm{r}}/r_{\rm{d}})_1$ dominates the second one. 
The largest difference between the critical tidal radii of a factor 
of $\sim 2$ is for low-inclination orbits with respect to the orbital plane of DSO. }
\label{fig_ratio}
\end{figure}

The ratio $r_{\rm{H,r}}/r_{\rm{H,d}}$ acquires values of $(1.9,2.1)$ for eccentricities $e \in (1,0)$. Therefore retrograde orbits are expected to be stable for larger distances from the host star, approximately by the factor of two for low-inclination orbits. 
See Fig. \ref{fig_ratio} for the polar plot of the dependence of the ratio of tidal Hill stability 
radii $r_{\rm{H,r}}/r_{\rm{H,d}}$ on the inclination $i_{\rm{D}}$ between a putative circumstellar 
accretion disk and the orbital plane of DSO.

\section{Conclusions}
\label{sec:conclusions}

In our sensitive imaging spectroscopy data set we measured prominent line emission from the DSO.
and determined new orbital parameters based on data from February till September 2014.
The source appeared to be single lined at all times.
Before the peribothron we detected red-shifted Br$\gamma$ line emission (at 2700~km/s) but no blue-shifted emission above the noise level
at the position of SgrA* or upstream the presumed orbit. 
After the peribothron we detected blue-shifted Br$\gamma$ line emission (at $-$3320~km/s) but no red-shifted emission above the noise level
at the position of SgrA* or downstream the presumed orbit.
We find a Br$\gamma$-line full width at half maximum of $50\pm10$\AA ~before and $15\pm10$\AA 
~after the peribothron transit, i.e. no significant
line broadening with respect to last year is observed.
Such a broadening would be expected in case of significant tidal interaction.
This is a further indication for the fact that the DSO is spatially rather compact.

We show that for a 1 - 2~\solm ~embedded pre-main sequence star hot accretion streams close to the star possibly in combination with disk winds can fully 
account for the luminous observed Br$\gamma$ emission with line widths
covering the full range from about 200 $\kms$ to 700 $\kms$.
The accretion material and the surrounding shell/disk provide enough extinction to
explain the infrared colors of the DSO \citep{eck13}.
The resulting line profile can be asymmetric and skewed to one side
calling for precaution when using the line emission to derive orbital parameters.
Following the pre-main sequence evolutionary tracks of low- and intermediate-mass stars by \citet{Siess2000}
we find that after an initial phase of a few 10$^{5}$ years 
1-2~\solm ~stars can stay for a major portion of their
T~Tauri stage with a luminosity of less than 10~\solar \citep[see also][]{chen2014}.
This is consistent with a dust temperature of 450~K and 
a possible spectral decomposition of the NIR/MIR spectrum \citep{eck13} 
of the DSO using the M-band measurement by \citet{gil12}.
Higher stellar masses would not comply with this luminosity limit and are not required to explain
the Br$\gamma$ line widths.
An embedded pre-main sequence star can also explain the increase of the Br$\gamma$ line width
assuming that tidal stretching and perturbation of the envelope
lead to an enhancement of the velocity dispersion in the accretion stream onto the central star
as the DSO approaches the peribothron. 
An identification of the DSO with a dust embedded star also puts the interpretation of a common
history of the DSO/G2 and G1 at risk \citep{pfu14b}. 
Due to the higher mass (1-2~\solm ~instead of 3 earth masses) a very much higher drag force than the 
one provided by the small source size 
would be required to connect the DSO orbit to that of G1.

We also find that the NIR flaring activity of SgrA* has not shown any 
statistically significant increment.
This points at the fact that the DSO had not yet reached its peribothron before May 2014. 
Even if the source has a stellar core a major part of the enshrouding cloud may be
dissolved during the peribothron passage.
Therefore increased accretion activity of SgrA* may still be upcoming.
The SgrA*/DSO system can be looked upon as a binary system and the Roche lobe picture can be 
adopted in which the Lagrange point L1 between the two objects is of special importance if 
mass transfer between the two objects needs to be considered.
If the central star has around one solar mass the L1 will get very close 
($\sim$1 AU) and may allow the dominant part of the
gas and dust to transit into the SgrA* dominated Roche lobe. 
As a result the low mass stellar core may be even less luminous after the transit than the  matter in
its immediate vicinity (i.e. 1 AU) before the transit.
This will, however, be only for a very short time and it is not clear if the gas close to the star will
remain in the Roche lobe of the star after peribothron or not.
For higher mass stellar cores that were heavily extincted before peribothron most of the material closer
to the stars (i.e. a few AU) may be largely unaffected by the transition and the stellar core may
by even cleared from extincting material and brighter in the NIR-bands than before the peribothron .

In the near future it will become increasingly difficult to measure the strength and spatial extent of
the line emission on the blue side of the orbit.
This is due to a high velocity star that is moving into this field from the north-west.
It will then be followed by S2 going though peribothron 
around 2017.9$\pm$0.35 \citep{gil09, eis03}
and S0-102 around 2021.0$\pm$0.3 \citep{mey14a,mey14c}. 
Strong continuum contributions and residual line features in the stellar atmospheres may make sensitive
observations very difficult and time consuming.

\acknowledgments
The research leading to these results received funding from the
European Union Seventh Framework Program (FP7/2007-2013) under grant agreement n312789.
This work has been financially supported by the
Programme National Hautes Energies (PNHE).
This work was supported in part by the Deutsche Forschungsgemeinschaft
(DFG) via the Cologne Bonn Graduate School (BCGS), 
the Max Planck Society through 
the International Max Planck Research School (IMPRS) for Astronomy and 
Astrophysics, as well as 
special funds through the University of Cologne.
M. Zajacek, B. Shahzamanian, S. Smajic, A. Borkar
are members of the IMPRS.
Part of this work was supported by fruitful discussions
with members of the European Union funded COST Action MP0905: Black
Holes in a violent Universe and 
the Czech Science Foundation -- DFG collaboration (No. 13-00070J).
This work was co-funded under the Marie Curie Actions of the European
Commission (FP7-COFUND).
Macarena Garc\'{\i}a-Mar\'{\i}n is supported by the German 
federal department for education and research (BMBF) under 
the project number 50OS1101.
We are grateful to all members
of the ESO PARANAL team.

\bibliographystyle{apj}
\bibliography{xcombined}

\end{document}